\documentclass[pdflatex,sn-nature]{sn-jnl}

\usepackage{nicefrac}%
\usepackage{upgreek}%
\usepackage{graphicx}%
\usepackage{multirow}%
\usepackage{amsmath,amssymb,amsfonts}%
\usepackage{amsthm}%
\usepackage{mathrsfs}%
\usepackage[title]{appendix}%
\usepackage{xcolor}%
\usepackage{textcomp}%
\usepackage{manyfoot}%
\usepackage{booktabs}%
\usepackage{algorithm}%
\usepackage{algorithmicx}%
\usepackage{algpseudocode}%
\usepackage{listings}%
\usepackage[mathscr]{euscript}

\usepackage[mathscr]{euscript}

\usepackage{geometry}

\begin{document}

\title[Article Title]{Coherent spin waves in a maximal entropy phase}

\author*[1]{ArnauRomaguera} \email{arnau.romaguera-camps@psi.ch}
\author[1]{Eugenio Paris} \email{eugenio.paris@psi.ch}
\author[1]{Elizabeth Skoropata} \email{elizabeth.skoropata@psi.ch}
\author[2]{Stefano Agrestini}  \email{stefano.agrestini@diamond.ac.uk}
\author[2]{Mirian Garcia-Fernandez} \email{mirian.garcia@diamond.ac.uk}
\author[3]{Marisa Medarde} \email{marisa.medarde@psi.ch}
\author[4]{Noah Schnitzer} \email{nis29@cornell.edu}
\author[5]{Lopa Bhatt} \email{lb628@cornell.edu}
\author[5,6]{Berit H. Goodge} \email{bhg37@cornell.edu}
\author[7,8,9]{Yun Yen} \email{yen@uni-bremen.de}
\author[7]{Matthias Krack} \email{matthias.krack@psi.ch}
\author[7,10]{Michael Schüler} \email{michael.schueler@psi.ch}
\author[3]{Romain Sibille} \email{romain.sibille@psi.ch}
\author[3]{Tom Fennell} \email{tom.fennell@psi.ch}
\author[3]{Daniel G. Mazzone}\email{daniel.mazzone@psi.ch}
\author[3]{Jakob Lass}\email{jakob.lass@psi.ch}
\author[8, 11, 12]{Ellen Fogh}\email{ellen.fogh@epfl.ch}
\author[13]{Anirudha Ghosh}\email{anirudha.ghosh@maxiv.lu.se}
\author[13]{Marco Caputo}\email{mcaputo@anybotics.com}
\author[1]{Carlos William Galdino}\email{carlos.galdino@diamond.ac.uk}
\author[1,8]{Zhijia Zhang}\email{zhijia.zhang@psi.ch}
\author[1]{Thorsten Schmitt}\email{thorsten.schmitt@psi.ch}
\author[1]{Milan Radovic}\email{milan.radovic@psi.ch}
\author[1]{Luc Patthey}\email{luc.patthey@psi.ch}
\author[1]{Hiroki Ueda}\email{hiroki.ueda@psi.ch}
\author[3,14]{Monica Ciomaga~Hatnean}\email{monica.ciomaga@psi.ch}
\author*[1]{Elia Razzoli}\email{elia.razzoli@psi.ch}

\affil[1]{PSI Center for Photon Science, Paul Scherrer Institute, Villigen-PSI, 5232, Switzerland}
\affil[2]{Diamond Light Source, Harwell Campus, Didcot, OX11 0DE, UK}
\affil[3]{PSI Center for Neutron and Muon Sciences, Paul Scherrer Institute, Villigen-PSI, 5232, Switzerland}
\affil[4]{Department of Materials Science and Engineering, Cornell University, Ithaca NY, 14853, USA}
\affil[5]{School of Applied and Engineering Physics, Cornell University, Ithaca NY, 14853, USA}
\affil[6]{Max Planck Institute for Chemical Physics of Solids, Dresden, 01187, Germany}
\affil[7]{PSI Center for Scientific Computing, Theory and Data, Paul Scherrer Institute, Villigen-PSI, 5232, Switzerland}
\affil[8]{École Polytechnique Fédérale de Lausanne (EPFL), Institute of Physics (IPHYS), Lausanne, 1015, Switzerland}
\affil[9]{Institute for Theoretical Physics and Bremen Center for Computational Materials Science, University of Bremen, Bremen, 28359, Germany}
\affil[10]{Department of Physics, University of Fribourg, Fribourg, 1700, Switzerland}
\affil[11]{Department of Physics, Technical University of Munich,   Garching b. Munich, 85748, Germany}
\affil[12]{Center for Quantum Engineering (ZQE), Technical University of Munich, Garching b. Munich, 85748, Germany}
\affil[13]{MAX IV Laboratory, Lund University, P.O. Box 118, Lund, SE-22100, Sweden}
\affil[14]{Materials Discovery Laboratory, Department of Materials, ETH Zürich, Zürich, 8093, Switzerland}
\abstract{
In solids, disorder is conventionally regarded as detrimental to coherence. It typically localizes and dampens collective excitations, as exemplified by Anderson localization or the broadening of magnetic modes in systems lacking long-range order. While high-entropy materials are specifically designed to harness disorder and stabilize homogeneous mixed-phase structures that can display unique properties, this same disorder is nonetheless expected to preclude the formation of coherent magnetic excitations.
To test the limits of this picture, we selected the antiferromagnetic system YBaCuFeO$_5$, as it features two distinct transition metal atoms with significantly different magnetic moments, rendering its spin dynamics exceptionally sensitive to local atomic ordering.
Combining resonant inelastic x-ray scattering and  linear spin wave theory, we reveal a surprising paradox: YBaCuFeO$_5$ exhibits an unexpected, entropy-driven mixed phase, in which disorder, rather than reducing the lifetime of the collective excitations, favors coherence. In this mixed phase, the spin waves remain dispersive, markedly distinct from those expected for an ordered ground state, and exhibit well-defined acoustic and optical branches separated by a large optical gap. These results demonstrate that in entropy-stabilized magnets, disorder can favor coherent collective modes previously thought to be exclusive to low-entropy systems.}

%%\keywords{keyword1, Keyword2, Keyword3, Keyword4}

\maketitle

High-entropy materials (HEMs)  are novel systems that take advantage of disorder to enable access to unique properties~\cite{Rost2015}. Stabilized  by high configurational entropy, HEMs feature multiple elements randomly occupying a single sublattice, forming  a single-phase multi-component material, termed a mixed state. Due to the high level of configurational complexity, emergent properties arise that cannot be attributed to their individual constituents, including exceptional thermal and mechanical stability~\cite{Aamlid2023}.
Although various chemical properties of HEMs have been well established, the distinctive factors that govern their magnetic behavior are not yet fully understood and remain the focus of intensive investigation~\cite{Sarkar2021}.
Quite generally, even low levels of atomic disorder, typically induced by dilute chemical substitution, can have large effects on the properties of magnetic materials. Long-range magnetic order can be suppressed while, in the presence of strong exchange couplings, short range correlations survive. For instance, in cuprate superconductors, spin waves can preserve some dispersion but are typically overdamped~\cite{LeTacon2011, Wang2022}. In other cases, magnon localization can be induced, and as a result flat magnon branches are observed~\cite{Lyo1972, Buyers1972, Tseng2022}.

Here, we employ resonant inelastic x-ray scattering (RIXS) to map the spin excitation spectrum of YBaCuFeO$_5$ (YBCFO), a material hosting both commensurate (CM) and incommensurate (ICM) antiferromagnetic (AFM) ordering~\cite{Morin2015, Morin2016, Lai2017, Shang2018, Romaguera2022}. Our results demonstrate that YBCFO  is in an entropy-driven mixed state and hosts coherent spin waves with a large  optical gap. Comparison with linear spin wave theory (LSWT) and atomistic spin dynamics (ASD) calculations  reveals that the dispersion in the mixed phase is very different from that of the low-entropy ground state. Instead, it is underdamped and almost indistinguishable from that predicted for  an ordered high-energy configuration. 

\begin{figure}[h]
    \centering
    \includegraphics[width=\textwidth]{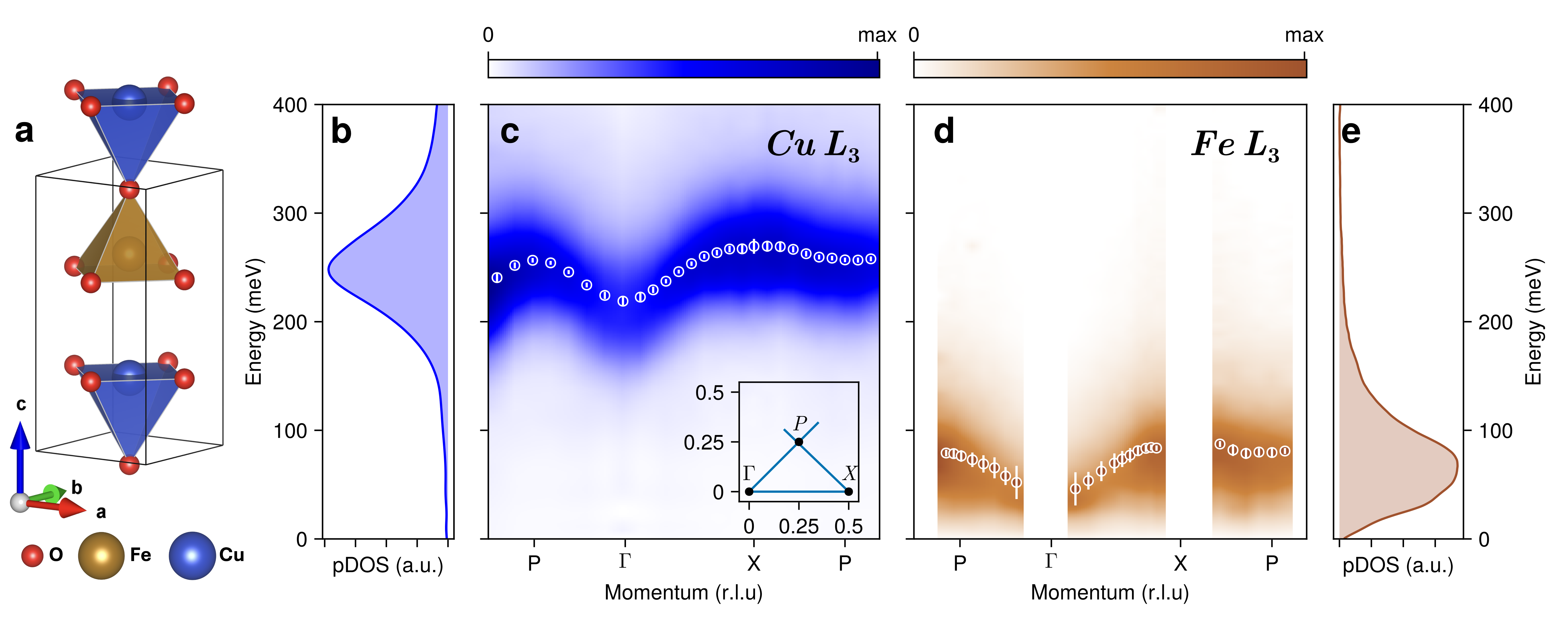}
    \caption{\textbf{RIXS measurements at Fe and Cu $L_3$ edges at 20~K.}
        \textbf{a}, Crystal structure of YBCFO generated using VESTA~\cite{momma2011vesta}, with the orientation of the tetragonal unit-cell axes indicated (Cu in blue, Fe in brown).
        \textbf{b}, Experimental RIXS intensity at Cu $L_3$-edge, integrated along high-symmetry path shown in the inset of \textbf{c}.
        \textbf{c, d}, Experimental $\mathscr{S}(\mathbf{q}, \omega)$ at Cu and Fe $L_3$ edges, respectively, after correction for self-absorption and the geometric spin-flip factor. Circles mark the extracted dispersion. 
        \textbf{e}, RIXS intensity at Fe $L_3$-edge, integrated along the same high-symmetry path.}
    \label{FigRIXS}
\end{figure}

YBCFO is a layered compound whose main structural units are CuO$_5$ and FeO$_5$ square pyramids that share apical oxygen atoms, forming bipyramidal units along the $c$ axis (see Fig.~\ref{FigRIXS}\textbf{a}). The single-crystal investigated exhibits CM antiferromagnetic order below $T_{\mathrm{CM}}=450$~K. Upon cooling below $T_S=180$~K, it undergoes a transition to an ICM spiral phase in which the in-plane spin order remains commensurate and the spin canting propagates along the $c$-axis~(see methods). 
For the CM phase, the magnetic structure is consistent with the combination of in-plane AFM couplings ($J^{ab}_{CuCu}$, $J^{ab}_{FeFe}$) together with alternating out-of-plane FM $J^{\perp}_{intra}$ (for the Cu--O--Fe bond, i.e., within a bipyramid) and AFM couplings $J^{\perp}_{inter}$ (between bipyramid), as obtained from density functional theory (DFT)~\cite{Morin2015}. In the spiral phase, low-density Fe--O--Fe impurity bonds within the bipyramids are believed to generate the FM frustration required to stabilize the ICM order at low temperature~\cite{Scaramucci2018}. 
In both phases, the Fe/Cu arrangement is crucial: the magnetic order requires an overall majority of the Cu--O--Fe bipyramids along $c$, with only a small fraction of Fe--O--Fe (and corresponding Cu--O--Cu) impurity bonds. By contrast, comparatively little attention has been devoted to the study of the Fe/Cu ordering within the $ab$ plane, where most previous work has assumed a stacking of monoatomic Fe and Cu 2D layers as the lowest-energy configuration~\cite{Morin2015, Dibyendu2018}.

RIXS spectra in the low-energy-loss region, collected at the Cu and Fe $L_3$ absorption edges,   are shown in Fig.~\ref{FigRIXS}, as a function of momentum transfer. 
The elastic and a non-dispersive phonon peak are subtracted from the data (see methods). Data are corrected for self-absorption~\cite{Robarts2021,Minola2015} and for the geometric spin-flip factor~\cite{Ament2009, Jia2016}. This allows a direct comparison of RIXS intensities to the atomically projected dynamical spin structure factors $\mathscr{S}_{Cu}(\mathbf{q}, \omega)$ and $\mathscr{S}_{Fe}(\mathbf{q}, \omega)$\textemdash the spin-spin correlation functions for Cu and Fe spin operators, respectively~\cite{Haverkort2010}.  Experimental undamped frequencies superimposed to the intensity plots in Fig.~\ref{FigRIXS}\textbf{c},\textbf{d} are obtained by fitting the data to a damped harmonic oscillator model (DHO)~\cite{Lamsal2016, Monney2016, Peng2018}.
At Fe $L_3$, the observed energy-momentum dispersion is in qualitative agreement with the expected behavior of a 2D square-lattice AFM. At zone boundary the magnon dispersion reaches its maximum, while $\omega(\mathbf{q})$ decreases linearly towards ${\Gamma}=(0, 0)$.
Quite surprisingly, at Cu $L_3$  the observed energy-momentum dispersion  has a large gap $\Delta \approx 218$~meV at ${\Gamma}$, and reaches the maximal energy of $\approx 269$ and $\approx 257$ meV at the high symmetry points ${X}=(0.5, 0)$ and ${P}=(0.25, 0.25)$, respectively.
No significant dispersion changes are observed across $T_S$ (see methods), as expected from the preserved in-plane CM order and the limited RIXS sensitivity to the small modifications of the out-of-plane dispersion between the spiral and CM phases~(see Supplementary Information~5).
The optical spin gap observed in the magnon spectra is among the largest ever measured, a rare occurrence since large gaps are uncommon even in materials with complex magnetic order. To better understand the nature of magnetic excitations and the origin of the large optical gap, we performed LSWT calculations for the expected CM magnetic structure.

\begin{figure}[h]
    \centering
    \includegraphics[width=\textwidth]{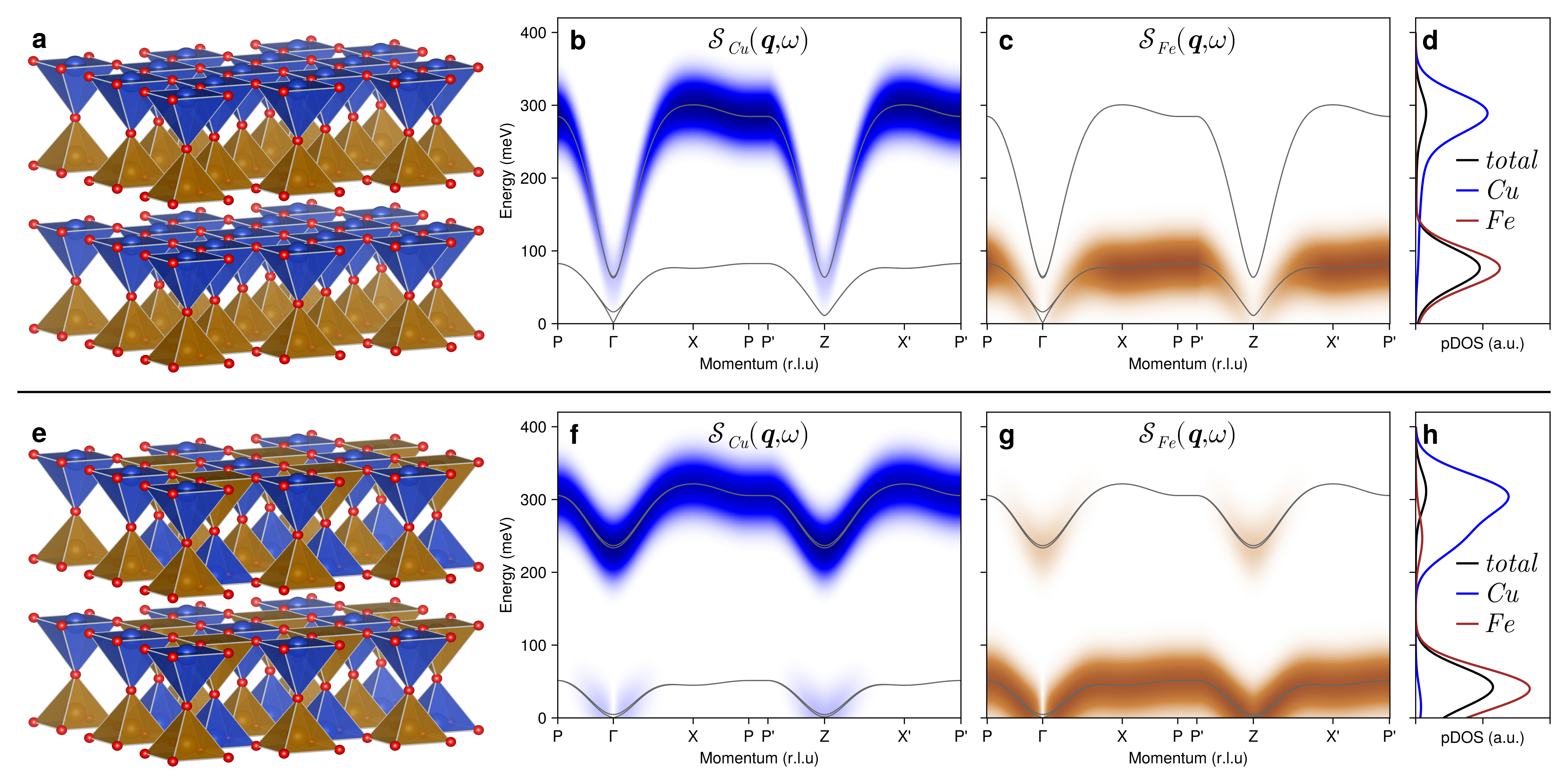}
    \caption{\textbf{LSWT calculations for two atomic-ordering configurations in YBCFO.} 
        \textbf{a}, Atomic configuration containing staggered monoatomic Fe and Cu planes (Cu in blue, Fe in brown).  
        \textbf{b--d}, LSWT results for the configuration in \textbf{a}, showing the Cu- and Fe- projected $\mathscr{S}(q,\omega)$, respectively, along high symmetry lines connecting $\Gamma = (0, 0, 0)$, $\text{P}=(0.25,0.25,0)$,  $\text{X}=(0.5,0,0)$, $\text{Z}  = (0, 0, 0.5)$, $\text{P}{'}=(0.25,0.25,0.5)$ and $\text{X'}=(0.5,0,0.5)$ (r.l.u. in units of $2\pi/a$, $2\pi/b$, $2\pi/c$). Thin gray lines indicate the LSWT eigenvalues.
        \textbf{e}, Atomic configuration with diatomic Fe and Cu planes in a checkerboard arrangement.         
        \textbf{f--h}, LSWT results for the configuration in \textbf{e}. 
        %Blue (brown) intensity plots in \textbf{f} and \textbf{g} show the atomic projected $\mathscr{S}(q,\omega)$ on Cu (Fe) atoms. 
        \textbf{d,h}, Total and atomically projected        dynamical spin structure factors integrated along the paths indicated in \textbf{b},\textbf{c} and \textbf{f},\textbf{g}. 
        Exchange parameters are obtained from DFT+U calculations~(see Methods).}
    \label{FigLSWT}
\end{figure}

In Fig.~\ref{FigLSWT}\textbf{a} we show the DFT lowest energy configuration of the crystal structure of YBCFO, which is consistent with the observed crystal symmetry.  In  this configuration, each 2D-plane is composed of a single atomic type (either Fe or Cu), with alternating  stacking along the $c$ axis.  Within each layer we expect a quasi-2D spin-wave dispersion that follows the characteristic dispersion of magnons in monoatomic square lattice.
Considering only nearest-neighbor (NN) interaction, we expect an energy at the zone boundary  to be of the order of $4J^{ab}_{CuCu}S_{Cu}\approx260$ meV and $4J^{ab}_{FeFe}S_{Fe}\approx90$ meV, for Cu and Fe layers, respectively. 
The degeneracy at $\Gamma$ is lifted by inter-layer couplings ($J^{\perp}_{inter}\approx-J^{\perp}_{intra} \equiv J^{\perp}$), giving rise to a small gap between optical and acoustic branches. The numerical results for the magnon dispersion $\omega(\mathbf{q})$ (gray lines) and $\mathscr{S}_{Cu/Fe}(\mathbf{q}, \omega)$ (blue/brown  intensity maps) in \textbf{b}, \textbf{c} confirm this picture, and  our analytic calculations show that the optical gap is given by $\sim 4\sqrt{S_{Fe}S_{Cu}J^\perp J^{ab}_{CuCu}} \approx 65$ meV (see Methods). 
Although the couplings may be slightly underestimated, an increase of $J^\perp$ and/or $J^{ab}_{CuCu}$ would still fail in reproducing the experimental dispersion. For instance, for a $J^{\perp}\approx 16$ meV, a value roughly ten times larger than calculated in DFT, the gap is of the order of 200 meV. Yet, the dispersion at $X$ would be renormalized by $\sim \sqrt{1+(S_{Fe}/S_{Cu})  (J^{\perp}/J^{ab}_{Cu})}\approx 1.3$, yielding $\omega (X)=330$ meV, which is inconsistent with the measured spectrum.  
Thus, while LSWT correctly predicts an optical branch predominantly visible at Cu $L_3$ edge, and an acoustic branch visible at Fe $L_3$ edge, it fails to reproduce the size of the observed optical gap and we need to look for a different explanation for its origin.

Sizable optical gaps ($\gtrsim 100$ meV) are typically the result of complex terms in the Hamiltonian, such as the pseudodipolar interaction~\cite{Kim2012}, the ferromagnetic Kitaev interaction~\cite{Chun2021}, the large inter-layer exchange coupling~\cite{Reznik1996} and the interaction of spin and itinerant carriers~\cite{Paris2025}. None of these interactions is predicted to reach significant magnitudes in YBCFO~\cite{Dey2018}.  
DFT calculations  show that atomic configurations that preserve the Cu--O--Fe bonds along $c$, but with a different atomic ordering in the $ab$ plane have a relatively small energy separation~\cite{Morin2015}. Given the high sensitivity of spin waves to the spatial coordination of the magnetic moments of the atoms in the unit cell, we need to consider the different ordering as a possible origin of the optical gap. 
LSWT calculations for a representative configuration -  chosen for its small unit cell -  are shown in Fig.~\ref{FigLSWT}\textbf{e}--\textbf{h}, while similar conclusions can be drawn for other configurations with similar energy. In this case, each 2D-plane contains both Cu and Fe atoms, which are ordered in a checkerboard fashion.  The magnon dispersion is that of weakly coupled diatomic square-lattice planes, and acoustic and optical branches are expected. 
The magnon dispersion $\omega(\mathbf{q})$ (gray lines) in Fig.~\ref{FigLSWT}\textbf{f}, \textbf{g} features a large optical gap. 
In the small  $J^\perp$ limit, the optical gap at $\Gamma$ is:

\begin{align}
    \label{eq1}
    \Delta(\Gamma) &= 4dS J^{ab}_{CuFe} \sqrt{1+(1+\frac{S_{Fe}S_{Cu}}{dS^2})\frac{J^\perp}{J^{ab}_{CuFe}}}
\end{align}
with $dS=S_{Fe} - S_{Cu}$.
In this model a gap $\Delta_{2D}(\Gamma) = 4dS J^{ab}_{CuFe}$ opens already in the 2D limit ($J^\perp=0$),  with small correction due to $J^\perp$  ($\Delta(\Gamma)= 1.03\Delta_{2D}(\Gamma)$ for $J^\perp\approx1.6$ meV). 
As shown in Figs.~\ref{FigLSWT}\textbf{f} and \textbf{g}, the experimental dispersion aligns more closely with the magnetic mode of the configuration featuring diatomic checkerboard-ordered 2D planes.
In this model, the gap originates from the difference in the exchange field between the two sublattices, i.e.  $\Delta = z_AJ^{ab}_{CuFe}S_A - z_BJ^{ab}_{CuFe}S_B= 4J^{ab}_{CuFe}dS$, with $z_i=4$ the number of nearest neighbors of site $i$.  The large value of the gap in YBCFO is thus due to the rare combination of large values for the coordination $z^i$, the exchange coupling $J^{ab}_{CuFe}$ and the difference in the spin quantum number of the two sites $dS$. For instance, in YIG a spin gap opens via the same mechanism. Here, the exchange field difference arises from the imbalance in the number of up and down spins in the two sublattices ($z^A=6$, $z^B=4$, $S_A=S_B=2.5$ and $J\approx7$ meV)~\cite{Princep2017}, yielding $\Delta = 6JS_A - 4JS_B= 2JS \approx 35$ meV, a much smaller value than in YBCFO.
Although the 2D-checkerboard model captures the main feature of the experiment, it does exhibit several limitations. First, the crystallographic unit cell in the checkerboard configuration is  $\sqrt{2} \times \sqrt{2}$ larger in the $ab$-plane than that of the 2D-monoatomic plane model. As such, we would expect to observe multiple extra peaks, e.g., the $(0.5, 0.5, 1)$ reflection should become observable, while it is absent in both our~(see Methods) and previous measurements~\cite{Lai2024, Romaguera2024}. 
Second, in the checkerboard ordered planes the magnetic moment is not compensated, each plane is ferrimagnetic and new magnetic reflections acquire finite intensity. Our LSWT predicts a large intensity for the magnetic peak (0, 0, 1.5) that is not present in our elastic neutron scattering  measurements~(see Methods).

\begin{figure}[h]
    \centering
    \includegraphics[width=\textwidth]{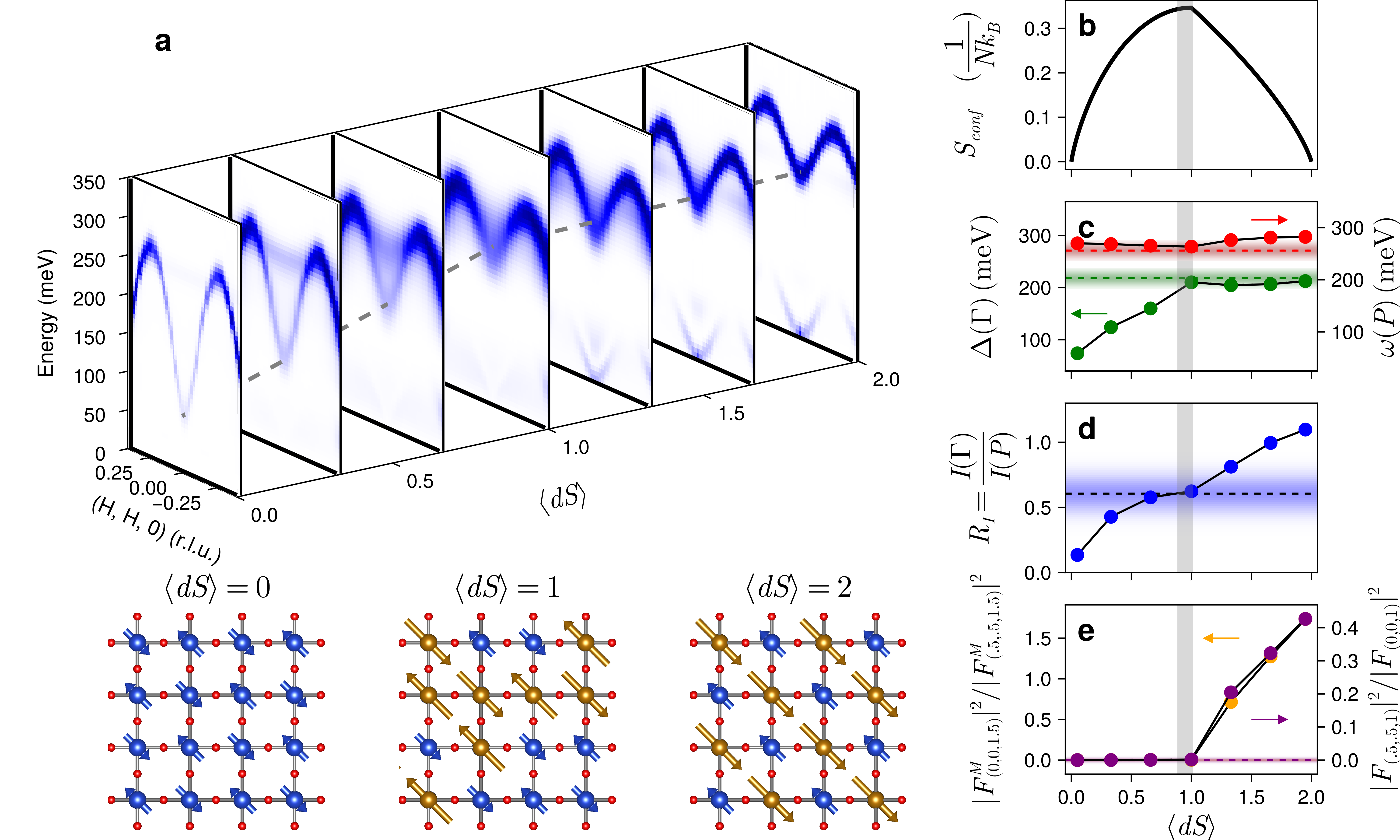}
    \caption{\textbf{Entropy dependence of the magnon optical-branch dispersion in YBCFO.}
    \textbf{a}, ASD $\mathscr{S}_{Cu}(\mathbf{q}, \omega)$ calculations as a function of the local ordering in the 2D planes $\langle dS \rangle$. Representative configurations of the 2D plane ordering for the ordered ($\langle dS \rangle=0, 2$) and high entropy ($\langle dS \rangle=1$) phases are shown at the bottom, as schematics of the atomic and spin configurations (Cu in blue, Fe in brown).  
    \textbf{b--e}, Properties of YBCFO as a function of $\langle dS \rangle$: Configuration entropy $S_{conf}$ (\textbf{b}), spin optical gap at $\Gamma$ and spin-waves energy at $P$ (\textbf{c}), normalized intensity of the optical branch (\textbf{d}), and lattice and magnetic Bragg reflection intensities (\textbf{e}). 
    Dashed lines in \textbf{c--e} represent the experimental values. Shaded areas correspond to Gaussian distributions with the FWHM fixed to the estimated experimental uncertainties. The gray shaded region indicates $\langle dS \rangle$ values for which $S_{conf}>.99 \max(S_{conf})$, that is, where the entropy exceeds $99\%$ of its maximum value. $S_{conf}$ is calculated for bilayers that conserve the Cu--O--Fe bipyramidal units. }
    \label{FigEntropy}
\end{figure}

These results indicate that the difference in the spin quantum numbers $dS$, for two in-plane NN atoms, can give rise to a large optical gap, already in a 2D model.  The lack of the additional diffraction peaks suggests that the system should lack the long-range checkerboard atomic order while maintaining a large local non-zero average  of $dS$. Long-range magnetic ordering should also be preserved, given the presence of main magnetic reflections. 

The  nature of this state is revealed by analyzing the expected spin-wave spectra for atomic configurations with different values of the $\langle dS \rangle = |\langle S_i - 1/4 \sum_{j \in NN(i) }S_j \rangle|$, the value of $dS$ mediated over the in-plane NNs in presence of atomic disorder. %Calculations are performed within the ASD framework, to allow for efficient scaling of the super cell size. 
 To obtain such spectra, we  perform ASD calculations using the stochastic Landau-Lifshitz dynamics~\cite{Dahlbom2025} for large unit cells, with different atomic distribution of Fe/Cu atoms.
 Given that the most dramatic changes induced by the different Fe/Cu ordering in Fig.~\ref{FigLSWT} involve the optical gap, we focus on the evolution of the optical branch in the following.
The results of the calculations as a function of $\langle dS \rangle$  for the optical branch, i.e., after projection on Cu atoms, are shown in Fig.~\ref{FigEntropy}\textbf{a}. The exchange couplings used in the model are given in Table~\ref{TableJs} and are very similar to the DFT values (see Methods).

\begin{table}[h]
\caption{Exchange coupling parameters used in the ASD calculations. $J>0$ indicates AFM coupling. Units are in meV.}\label{TableJs}%
\begin{tabular}{@{}lllllll@{}}
\toprule
$J^{ab}_{CuCu}$  & $J^{ab}_{FeFe}$  & $J^{ab}_{CuFe}$    &  $J^{\perp}_{inter}$   & $J^{\perp}_{intra}$ & $J^{'}_{CuCu}$  & $J^{'}_{FeFe}$ \\
\midrule
125 & 9 & 26 & 1.1 & -1.1 & -10 & -1.5  \\
\botrule
\end{tabular}
\end{table}

The results for mono-atomic $\langle dS \rangle=0$ and checkerboard ordered planes $\langle dS \rangle=2$ reproduce what is expected from Fig.~\ref{FigLSWT}. The optical gap does not smoothly interpolate between the two results, as one could expect from a phenomenological relation $\Delta(\Gamma) \approx 4 J^{ab}_{CuFe} \langle dS \rangle$. Instead, the gap quickly increases to reach  $\sim  215$ meV for $\langle dS \rangle=1$ and then saturates. 
The spin dynamic associated with this phase is almost identical to that of the checkerboard-ordered phase.  The $\langle dS \rangle=1$ state is a maximal configurational-entropy ($S_{conf}$) phase with equal Fe and Cu occupation in the 2D planes, and lack of long-range atomic order [see Fig.~\ref{FigEntropy}\textbf{b}]. In the context of HEMs, this phase is termed binary mixed phase, indicating the mixed equiatomic occupation of each site by two atomic species.  An entropy‑stabilized binary mixed phase has been reported for a spinel structure~\cite{NAVROTSKY1968}, but not in a perovskite system. Comparison between the model and the experimental results for given key properties of the spin-waves excitations further confirms that YBCFO is in a maximal entropy mixed phase. The spin gap and $\omega(P)$ dependence on $\langle dS \rangle$ are shown in Fig.~\ref{FigEntropy} \textbf{c}.  Both show good agreement with their experimental value (red and green dashed lines extracted from Fig.~\ref{FigRIXS}, respectively) for $\langle dS \rangle \approx 1$.
The intensity of the optical branch at the zone center, which we plot, normalized by its value at $P$, in panel \textbf{d}, is also very sensitive to changes in the atomic ordering. The calculated values monotonically increase from the low value expected for the monoatomic planes case, which would be vanishing in the two dimensional limit, and reaches the highest value in the checkerboard phase, with a magnitude very close to the expected value for a pure 2D-checkerboard phase (${R}_I= S_{Fe}/dS= 1.25$~, see Supplementary Information). The best agreement with the experimental value (black dashed line) and the model is obtained for the maximal entropy mixed phase, the $\langle dS \rangle =1$ phase at the peak of $S_{conf}$. We also show the evolution of the intensity of the extra-peaks in both lattice and magnetic Bragg reflections, which are non-negligible only for $\langle dS \rangle$ values above those of the mixed phase, as expected. We note here that, while previous structural refinements assuming preferential occupation of Cu and Fe onto distinct layers fit the data well, a fully disordered maximal-entropy model yields a comparable fit~\cite{Morin2015}. Structural refinement alone is thus insufficient to uniquely determine the in-plane Fe/Cu distribution due to the similar atomic form factor of Cu and Fe.

The origin of the maximal-entropy stabilization lies in the very high growth temperature and fast cooling rates employed for YBCFO. This procedure is typically employed in order to raise $T_{S}$~\cite{Morin2016}, by increasing the number of Fe--O--Fe impurity bonds~\cite{Scaramucci2018}.
Our results demonstrate that this procedure has the so-far unexplored effect of inducing a maximal atomic disorder in the $ab$ planes. Indeed, this is not an entirely unexpected result. Although configurations with a high density of Fe-O-Fe impurity bonds introduce a high enthalpy change, given their much higher  DFT energy ($>1$ eV), the energies of configurations including diatomic checkerboard ordered ab-planes are all within $90 - 200$ meV above the predicted ground state~\cite{Morin2015}. For the maximal entropy configuration we explored,  considering a $\Delta S_{conf}\approx 0.35$, $\Delta S_{imp}\approx 0.18$ - the entropy contribution of the dilute Fe--O--Fe impurities~(see Supplementary Information) - and a growth temperature of about $1300$ K, a phase with enthalpy $\Delta H\sim60$ meV above the ground state can be stabilized. Given that the contributions ignored here, e.g., the lattice deformation  entropy and growth kinematics, can further increase $\Delta H$, we estimate that a stabilization of the mixed phase is thermodynamically favorable. 

\begin{figure}[h]
    \centering
    \includegraphics[width=\textwidth]{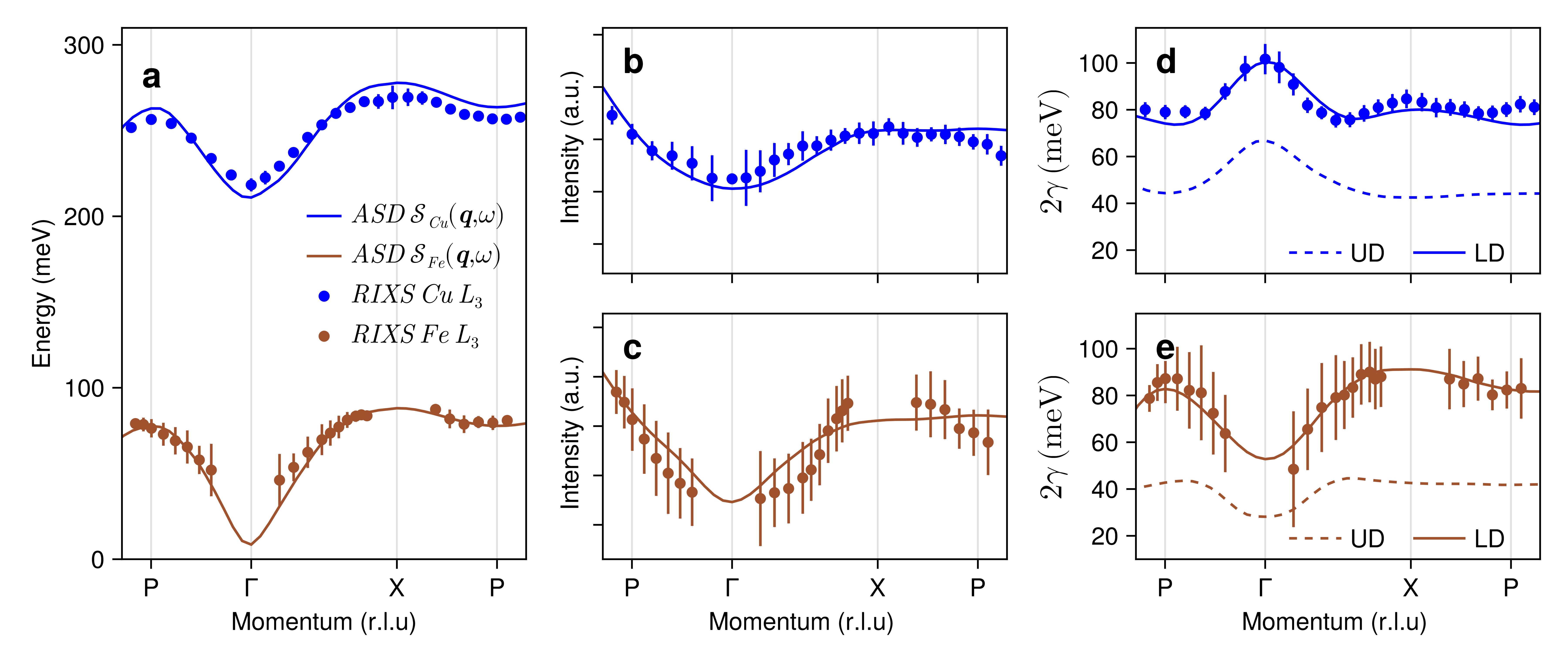}
    \caption{\textbf{Experimental and simulated momentum dependence of magnon peak parameters.}
    \textbf{a}, Undamped mode energies $\omega_0$ extracted from DHO fits to the experimental RIXS spectra, compared with the simulated peak positions for the high-entropy configuration.
    \textbf{b,c}, Integrated intensities obtained from numerical integration of the fitted DHO component for the Cu and Fe $L_3$ edges, respectively, compared with the simulated $\mathscr{S}(\mathbf{q},\omega)$ projected onto Cu and Fe atoms.
    \textbf{d,e}, Experimental linewidths $2\gamma$ (approximately the peak full width at half maximum) extracted from the DHO fits, compared with calculations for undamped (UD) and locally damped (LD) dynamics. %, and non-local damping (NLD) . 
    Blue and brown symbols denote values extracted from the experimental data at the Cu and Fe $L_3$ edges, respectively. All calculations include a finite instrumental energy-resolution broadening contribution.}
    \label{FigFWHM}
\end{figure}

Fig.~\ref{FigFWHM} shows that the mixed phase model ASD calculations accurately reproduce experimental dispersions, intensities, and damping. 
For the  optical branch, the experimental ratio is small, $\gamma/\omega\leq0.22$ across all studied $q$-points, indicating that the optical magnons are underdamped, and thus coherent~\cite{Lamsal2016, Monney2016, Peng2018}, which is striking given the significant influence of atomic disorder on the dispersion at $\Gamma$. In the acoustic branch, the experimental  ratio $\gamma/\omega$ is larger, but remains within the underdamped limit ($\gamma/\omega\sim 0.6 <1$).
To determine the origin of the anisotropy of $\gamma$, we compared a nearly undamped (UD) scenario ($\alpha^{Cu/Fe}_{i,i}=0.01$) with a local-damping (LD) case.
A peak at $\Gamma$ appears even in the UD calculation, signaling that it originates from the configurational disorder of the maximal entropy phase. For the acoustic branch, the effect of the local atomic disorder is more pronounced away from $\Gamma$. Finally, the inclusion of finite Gilbert damping  ($\alpha^{Cu}_{i,i}\sim 0.075$  and $\alpha^{Fe}_{i,i}\sim 0.26$)  in the LD calculations yields good agreement with the experimental data.

In conclusion, we have demonstrated that YBaCuFeO$_5$ hosts an entropy-driven mixed phase where atomic disorder facilitates, rather than destroys, coherent magnetic excitations. In contrast to other perovskite HEM systems, where no clear spin-wave dispersion is observed~\cite{zinouyeva2025, Zhang2019}, we report here an underdamped, and thus coherent, spin-wave dispersion. 
This coherence is sustained by the preserved long-range magnetic order and the near-invariance of the local exchange field ($J_{CuCu}S_{Cu} \sim J_{CuFe}S_{Fe}$, $J_{FeFe}S_{Fe} \sim J_{CuFe}S_{Cu}$), alongside the large optical gap that pushes $\omega^{opt}$ to very high energies~(see Supplementary Information). The large optical spin gap is driven by the profound difference in spin quantum number ($dS$) and the strong exchange coupling ($J_{FeCu}$) between the Fe and Cu atoms. 
Remarkably, the mixed phase retains the essential 
spin-waves characteristics of an unattainable ordered checkerboard phase, a configuration that cannot be thermodynamically stabilized due to its  high energy and low entropy.
Given that other high-spin transition-metal ions possess similar spin magnitudes and exchange strengths~\cite{Wuzong1993,Sharma2023}, we expect this mechanism to be potentially active in other similar disordered diatomic HEMs,  such as YBaCuCoO$_5$~\cite{Suescun2005}. Our results demonstrate that by strategically utilizing the contrast in spin magnitudes and the resulting exchange-field imbalance within high-entropy frameworks, it becomes possible to engineer disordered materials that host robust, well-defined, and high-energy collective modes, providing a new platform for studying emergent quantum effects in complex, disordered magnetic landscapes.

\backmatter

\clearpage

\bmhead{Acknowledgements}
The resonant inelastic x-ray scattering experiment was performed at the I21 beamline at the Diamond Light Source (proposal MM40202) and at Furka endstation at Athos beamline of SwissFEL. CWG, ZZ, and TS acknowledge funding from the Swiss National Science Foundation through project no. 207904. CWG acknowledges funding from the European Union’s Horizon 2020 research and innovation program under the Marie Sklodowska-Curie grant agreement No. 884104 (PSI-FELLOW-II-3i program). EF was funded by the European Research Council through the Synergy network HERO (Grant No. 810451) and by the Swiss National Science Foundation through Project Grant No. 188648.

\section*{Declarations}

\paragraph*{Funding:}
PSI Research Grant 2022 (AR, HU, MCH,  ER);

\paragraph*{Competing interests:}
The authors declare no competing interests.

\paragraph*{Data and materials availability:}
All data needed to evaluate the conclusions are available in the main text or the Supplementary Information.

\paragraph*{Author contributions:} Conceptualization: ER \\
Sample preparation: AR, MM and MCH \\
Methodology:  AR and ER\\
Investigation: AR, EP, ES, SA, NS, LB, BHG, YY, MK, MS, RS,
TF, DGM, JL, EF, AG, CWG, ZZ, TS, MR, HU and ER \\
Formal analysis: AR and ER \\
Visualization: AR and ER \\
Funding acquisition: HU, MCH and ER \\
Writing – original draft: AR and ER \\
Writing – review \& editing: all authors

\section*{Methods}\label{sec:methods}

\subsection*{Sample preparation and characterization}

A single crystal of YBaCuFeO$_5$ was grown by the traveling solvent floating-zone technique~\cite{Lai2015} using a laser-diode-heated floating zone furnace at the Materials Discovery Laboratory (ETH Zürich). Phase purity, stoichiometry, and crystallinity were characterized by energy-dispersive x-ray spectroscopy (EDX), powder and single-crystal x-ray diffraction, magnetic susceptibility, and single-crystal neutron diffraction.

DC magnetic susceptibility was measured under field-cooled (FC) conditions using a SQUID magnetometer (MPMS, Quantum Design), with an applied field of 0.2~T applied along the crystallographic $a$ and $c$ axes (Extended Data Fig.~\ref{Fig_Transitions}). Measurements above 300~K were performed using a high-temperature oven insert. Two anomalies are resolved in $\chi_a(T)$, corresponding to the paramagnetic-to-commensurate antiferromagnetic transition at $T_{\mathrm{CM}}=450$~K and the commensurate-to-spiral transition at $T_S=180$~K. By contrast, $\chi_c(T)$ shows no pronounced anomaly at $T_S$, consistent with moments confined to the $ab$ plane in the spiral phase and with previous reports on YBCFO single crystals~\cite{Romaguera2024,Lai2017}.

Single-crystal neutron diffraction measurements were performed on the ZEBRA diffractometer at the Swiss Spallation Neutron Source (SINQ, PSI) using a neutron wavelength of $\lambda=1.383$~\AA. Temperature-dependent $Q$ scans were collected along $(0.5,0.5,L)$ in the vicinity of $L=0.5$ (Extended Data Fig.~\ref{Fig_Transitions}\textbf{b}--\textbf{d}). Below $T_S$, two satellite reflections appear symmetrically about the commensurate position, confirming the onset of incommensurate spiral order with propagation vector $\mathbf{k}_{\mathrm{ICM}}=(1/2,1/2,1/2\pm q_S)$. The incommensurability increases continuously on cooling below $T_S$ and saturates at $q_S=0.095(5)$~r.l.u. at 10~K, consistent with a second-order commensurate-to-incommensurate transition. A residual commensurate contribution at $\mathbf{k}_{\mathrm{CM}}=(1/2,1/2,1/2)$ remains below $T_S$, with an integrated intensity of approximately 10\% of the total magnetic scattering, in agreement with previous single-crystal reports~\cite{Romaguera2024}.

Unless otherwise stated, YBCFO reciprocal space directions are indexed using the parent tetragonal crystallographic unit cell used throughout the literature, with lattice parameters $a=b$ and $c$ (see Fig.~\ref{FigRIXS}\textbf{a}). For the investigated sample, x-ray diffraction yields $a=b=3.87407(15)$~\AA\ and $c=7.66165(41)$~\AA\ at room temperature. Momentum transfer is expressed in reciprocal lattice units (r.l.u.) of this parent cell as $\mathbf{Q}=(h,k,l)$, with reciprocal lattice vectors $(2\pi/a,\,2\pi/b,\,2\pi/c)$.

\subsection*{Spectrometers and data processing}

RIXS measurements were performed at the I21 beamline of Diamond Light Source~\cite{Zhou2022}, and additional temperature-dependent RIXS measurements were collected at the Furka endstation of SwissFEL~\cite{furka_website}. Measurements were carried out on high-quality YBaCuFeO$_5$ single crystals mounted on a copper sample plate. A crystal with [001] surface orientation was used in order to access the relevant in-plane high-symmetry directions in reciprocal space. The samples are non-air-sensitive and therefore did not require cleaving in vacuum prior to measurement. The sample holder was mounted on the cryostat manipulator, and measurements were performed at low temperature ($T \approx 20$~K), corresponding to the spiral magnetic ground state.

RIXS spectra were collected at the Cu $L_3$ edge ($E_i = 931.76$~eV) and the Fe $L_3$ edge ($E_i = 709.5$~eV). At I21, the overall energy resolution was approximately 37~meV at the Cu $L_3$ edge and 28~meV at the Fe $L_3$ edge for the selected monochromator and spectrometer settings. For measurements at the Cu $L_3$ edge at Furka, the spectrometer yielded an overall energy resolution of approximately 200~meV. In all measurements discussed here, linear horizontal (LH) $\pi$-polarized incident x-rays were used with respect to the $a$-$c$ scattering plane.

The experimental geometry is illustrated in Extended Data Fig.~\ref{Fig_RIXS_Geometry}\textbf{a}. To access the directions along which the magnetic excitations are most dispersive, the momentum transfer was varied at fixed detector scattering angle $\Omega$ by rotating the sample incidence angle $\theta$ and the azimuthal angle $\phi$. For the [001]-oriented crystal, this geometry provides access to in-plane momentum cuts within the $[h,0,l]$ and $[h,h,l]$ scattering planes. The corresponding momentum-space trajectory in the projected $(h,k)$ plane is shown in Extended Data Fig.~\ref{Fig_RIXS_Geometry}\textbf{b}, where the measured path connects the high-symmetry points $P$, $\Gamma$, $X$, and $P$. To maximize momentum transfer, a near-backscattering geometry was used, with fixed detector angle $\Omega=150^\circ$ (and $\Omega=130^\circ$ at Furka). The sample angles $\theta$ and $\phi$ required to follow the selected trajectory are shown in Extended Data Fig.~\ref{Fig_RIXS_Geometry}\textbf{c},\textbf{d}, while the corresponding reciprocal-space coordinates $(h,k,l)$ are plotted in Extended Data Fig.~\ref{Fig_RIXS_Geometry}\textbf{e},\textbf{f} for the Cu $L_3$ and Fe $L_3$ measurements, respectively.

The momentum transfer $\mathbf{Q}=(h,k,l)$ was calculated from the sample incidence angle $\theta$, azimuthal angle $\phi$, incident photon energy $E_i$, and fixed detector angle $\Omega$. 
The sample alignment was optimized using the specular elastic signal. RIXS spectra were recorded on an area detector and reduced by integrating the detected photon counts along the non-dispersive detector direction after subtraction of a dark-image background and correction for detector curvature. The energy-per-pixel calibration and an initial zero of energy loss were determined from reference spectra acquired on amorphous carbon tape mounted next to the sample. For each spectrum, the zero-loss position was then further refined during the fitting procedure from the fitted center of the elastic peak. All spectra were normalized by the acquisition time.

\subsection*{Spectral fitting}

The low-energy part of each spectrum was modeled as the sum of an elastic contribution, a non-dispersive phonon contribution, a magnetic contribution, and a constant background. The elastic peak and the phonon feature near $E^{Fe}_{\mathrm{loss}}\approx 33$~meV and  $E^{Cu}_{\mathrm{loss}}\approx 58$~meV were described by pseudo-Voigt functions, while the magnetic contribution was modeled by a Bose-weighted damped harmonic oscillator (DHO) response,
\begin{equation}
    I_{\mathrm{mag}}(\mathbf{q},\omega)=A\,\chi''(\mathbf{q},\omega)\,[n(\omega)+1],
\label{eq:DHO}
\end{equation}
where $\omega \equiv E_{\mathrm{loss}}$, $A$ is an intensity prefactor, $n(\omega)=\left(e^{\hbar\omega/k_{\mathrm B}T}-1\right)^{-1}$ is the Bose occupation factor, and
\begin{equation}
    \chi''(\mathbf{q},\omega)=
    \frac{4\omega_0\gamma\omega}
    {(\omega^2-\omega_0^2)^2+(2\gamma\omega)^2}.
\end{equation}
Here $\omega_0$ is the undamped excitation energy and $\gamma$ is the damping parameter. In the weak-damping limit, the full width at half maximum of the magnetic peak is approximately $2\gamma$. The total fit function was therefore
\begin{equation}
    I(\omega)=I_{\mathrm{el}}(\omega)+I_{\mathrm{ph}}(\omega)+I_{\mathrm{mag}}(\omega)+C.
\end{equation}
At the Cu $L_3$ edge, an additional high-energy-loss shoulder was observed on the right side of the main magnon peak. This extra spectral weight, consistent with a bimagnon-related contribution, was modeled by including an additional peak component in the fit. 
Fits were performed independently at each momentum point using least-squares minimization with the \texttt{lmfit} Python package. The undamped excitation energies plotted in Fig.~\ref{FigFWHM}\textbf{a} were taken from the fitted parameter $\omega_0$, the intensities shown in Fig.~\ref{FigFWHM}\textbf{b},\textbf{c} were obtained by numerical integration of the fitted DHO component associated with the main magnon peak, and the linewidths shown in Fig.~\ref{FigFWHM}\textbf{d},\textbf{e} were taken as the fitted  $2\gamma$.

\subsection*{Intensity corrections}

The measured RIXS intensity is affected by both self-absorption inside the sample and by the geometry-dependent local spin-flip cross section. To enable a quantitative comparison of the momentum dependence of the magnetic spectral weight, we corrected the fitted magnon intensities for both effects, following the approach commonly used in $L_3$-edge magnetic RIXS analysis~\cite{Robarts2021}. The correction workflow and the quantities entering it are summarized in Extended Data Figs.~\ref{Fig_XAS_SA}, \ref{Fig_IntensityCorrections}, and \ref{Fig_Corrected-Spectra}. The corrected magnetic intensity is taken to scale as
\begin{equation}
    I_{\mathrm{corr}} \propto I_{\mathrm{meas}} \frac{C_{\mathrm{SA}}}{R_{\mathrm{spin}}},
\label{eq:Icorr}
\end{equation}
where $C_{\mathrm{SA}}$ is the self-absorption correction factor and $R_{\mathrm{spin}}$ is the single-ion spin-flip cross-section factor. This factorization is motivated by the standard approximation that the magnetic RIXS intensity can be written as the product of a local atomic spin-flip matrix element and the collective dynamical spin response~\cite{Haverkort2010}.

\paragraph*{Self-absorption correction.}
Self-absorption was calculated from x-ray absorption spectra measured on the same crystal at the I21 beamline. Orientation-dependent total-electron-yield (TEY) XAS were collected with the electric field predominantly aligned along the crystallographic $a$ and $c$ directions, providing the anisotropic absorption form factors $f_a(E)$ and $f_c(E)$ at the Cu and Fe $L_3$ edges (Extended Data Fig.~\ref{Fig_XAS_SA}\textbf{c},\textbf{g}). These XAS data were also used to select the incident energies for the RIXS measurements and to evaluate the absorption coefficients at both the incident photon energy $E_i$ and the emitted photon energy $E_f = E_i - E_{\mathrm{loss}}$.

The scattering geometries used to evaluate the absorption coefficients for the two main momentum planes in the RIXS measurements are shown in Extended Data Fig.~\ref{Fig_XAS_SA}\textbf{a},\textbf{b}. Following Ref.~\cite{Robarts2021}, the self-absorption factor was written as
\begin{equation}
    C_\mathrm{SA}(\theta,\Omega,\hat{\epsilon},\hat{\epsilon}')
    =
    \frac{
    \mu_i(\hat{\epsilon})\sin(\Omega-\theta)
    +
    \mu_f(\hat{\epsilon}')\sin\theta}
    {\sin(\Omega-\theta)},
\label{eq:CSA}
\end{equation}
where $\theta$ is the grazing incidence angle, $\Omega$ is the fixed detector scattering angle, and $\mu_i$ and $\mu_f$ are the polarization-dependent absorption coefficients for the incident and scattered photons, respectively. For $\pi$-polarized incident light, as used in the present work, the incident absorption coefficient is
\begin{equation}
    \mu_i^{\pi}=f_a(E_i)\sin^2\theta + f_c(E_i)\cos^2\theta.
\end{equation}
For the emitted photons, the corresponding coefficients were evaluated at the exit angle $\Omega-\theta$ and at the emitted energy $E_f$. For the non-spin-flip elastic and phonon channel we used the $\pi\rightarrow\pi'$ geometry,
\begin{equation}
    \mu_f^{\pi'}=f_a(E_f)\sin^2(\Omega-\theta)+f_c(E_f)\cos^2(\Omega-\theta),
\end{equation}
whereas for the magnetic channel we used the $\pi\rightarrow\sigma'$ geometry,
\begin{equation}
    \mu_f^{\sigma'}=f_a(E_f).
\end{equation}
This choice follows the standard approximation that the outgoing polarization is preserved for elastic and phonon scattering, while the single-magnon channel is dominated by polarization-flipped scattering.

The XAS-derived absorption form factors and the resulting self-absorption factors are shown in Extended Data Fig.~\ref{Fig_XAS_SA}. Panels~\textbf{d}/\textbf{e} and \textbf{h}/\textbf{i} show the calculated self-absorption factor $C_{\mathrm{SA}}$ for the elastic/phonon and magnetic channels at the Cu and Fe $L_3$ edges, respectively, as a function of energy loss and incidence angle $\theta$. Representative angular cuts at the selected emitted energies are shown in Extended Data Fig.~\ref{Fig_XAS_SA}\textbf{f},\textbf{j}. As expected, the correction is strongest at low energy loss and for large grazing-incidence angles, where re-absorption of the outgoing photons is maximal.

\paragraph*{Spin-flip cross-section correction.}
In addition to self-absorption, the magnetic RIXS intensity contains a geometry-dependent local spin-flip factor. To remove this contribution, we calculated a single-ion spin-flip cross section $R_{\mathrm{spin}}(\theta)$ using \textsc{EDRIXS}~\cite{EDRIXS} single-ion multiplet calculations for Cu$^{2+}$ ($3d^9$) and Fe$^{3+}$ ($3d^5$) in the local crystal-field environments appropriate for YBCFO. The local ordered moment was used as the quantization axis, and a small Zeeman field was included to lift the spin degeneracy of the ground state. Since the measurements were performed in the spiral phase, the local moment direction varies continuously along the helix. Accordingly, the spin-flip cross section was evaluated for multiple local-field orientations sampling a full $2\pi$ rotation of the spiral, and the final correction factor was obtained by averaging over these orientations. The resulting orientation-averaged integrated spin-flip spectral weight as a function of $\theta$ defines $R_{\mathrm{spin}}^{\mathrm{Cu}}(\theta)$ and $R_{\mathrm{spin}}^{\mathrm{Fe}}(\theta)$. 

Equation~(\ref{eq:Icorr}) was then applied to the integral magnon peak intensities extracted from the DHO fits. The effect of the combined correction on the intensities is shown in Extended Data Fig.~\ref{Fig_IntensityCorrections}. Panels~\textbf{a},\textbf{b} compare the measured and corrected integrated intensities, while panels~\textbf{c},\textbf{d} show separately the magnetic self-absorption factor $C_{\mathrm{SA}}^{\mathrm{mag}}$, the spin-flip factor $R_{\mathrm{spin}}$, and the total correction factor $C_{\mathrm{SA}}^{\mathrm{mag}}/R_{\mathrm{spin}}$ along the measured momentum trajectory. For comparison across momentum, the correction factors were normalized to unity at $\Gamma$. 

\paragraph*{Correction of the displayed spectra.}
For the display of corrected spectra, shown in Extended Data Fig.~\ref{Fig_Corrected-Spectra} and used to generate the color maps in Fig.~\ref{FigRIXS}, we applied an energy-dependent effective correction factor constructed from the fitted spectral decomposition,
\begin{equation}
    C_\mathrm{eff}(\omega)
    =
    w_\mathrm{el}(\omega)\,C_\mathrm{SA}^{\mathrm{el}}(\omega)
    +
    w_\mathrm{mag}(\omega)\,\frac{C_\mathrm{SA}^{\mathrm{mag}}(\omega)}{R_\mathrm{spin}},
\label{eq:Ceff}
\end{equation}
where $w_\mathrm{el}(\omega)$ and $w_\mathrm{mag}(\omega)=1-w_\mathrm{el}(\omega)$ are the fractional elastic/phonon and magnetic contributions at each energy loss, as obtained from the spectral fit. The corrected spectrum was then computed as
\begin{equation}
    I_\mathrm{corr}(\omega)=\left[I_\mathrm{meas}(\omega)-I_\mathrm{bg}\right]\,C_\mathrm{eff}(\omega),
\end{equation}
where $I_\mathrm{bg}$ is the fitted constant background. Extended Data Fig.~\ref{Fig_Corrected-Spectra} compares representative uncorrected and corrected spectral stacks for the Cu and Fe $L_3$ edges. For the 2D color maps in Fig.~\ref{FigRIXS}\textbf{c},\textbf{e}, the fitted elastic and phonon contributions were additionally subtracted after applying the correction in order to emphasize the magnetic excitations.

\subsection*{Temperature dependence of the optical branch}

To probe the temperature dependence of the optical magnon branch across the commensurate-to-spiral at $T_S=180$~K, we performed additional Cu $L_3$-edge RIXS measurements at the Furka endstation of SwissFEL on the same YBCFO crystal. Spectra were collected at 200~K (commensurate phase) and 19~K (spiral phase) at the in-plane momenta $\mathbf{Q}=(0,0)$ and $\mathbf{Q}=(0.34,0)$~r.l.u., corresponding to the zone center and a representative momentum along the $\Gamma$--$X$ path (Extended Data Fig.~\ref{fig_Tdep_Furka}). 

Owing to the lower energy resolution of the Furka spectrometer, and for clarity, the elastic contribution was subtracted from the displayed spectra. Within experimental uncertainty, no significant changes are observed in the peak energy (optical gap and bandwidth) or in the overall lineshape between the two temperatures. This confirms that the optical branch is essentially unaffected across $T_S$, consistent with the preserved in-plane magnetic order and the small changes in out-of-plane dispersion discussed in the main text and Supplementary Information~5.

\subsection*{Simulations}

\paragraph*{DFT+U calculations}

The exchange coupling used for the LSWT simulations are taken from our DFT+U numerical calculations. The electronic structure calculations were performed using the Vienna Ab initio Simulation Package (VASP), version 6~\cite{Kresse1996}. The interaction between the core and valence electrons was described using the Projector Augmented-Wave (PAW) method~\cite{Kresse1999}. To account for the strong on-site Coulomb interactions of the Fe and Cu $3d$ electrons, the LSDA+U approach~\cite{PZ1981} was employed using the rotationally invariant formulation proposed by Dudarev et al.~\cite{Dudarev1998} with the effective $U_\text{eff} = U - J$ values of 4~eV and 8~eV for Fe and Cu, respectively. A $\sqrt{2}\times \sqrt{2}\times 2$ supercell of YBaCuFeO$_5$ with a $\Gamma$-centered  $8\times 8 \times 4$ $k$-point mesh and a plane-wave cutoff of 600~Ry served as model system for the density functional theory (DFT) calculations. The atomic positions were relaxed within the detected space group symmetry while the lattice constants were kept fixed at their experimental values. The $J$ coupling constants were calculated as described in Morin et al.~\cite{Morin2015}. In-plane (out-of-plane)  nearest-neighbor couplings are denoted by $J^{ab}$ ($J^c$), while the coupling bond is denoted in the subscript.  

For checkerboard planes configuration we also calculated the next-nearest-neighbors inplane interaction $J'$ by constructing a large ($2\times 2\times 2$) supercell with a $\Gamma$-centered  $6\times 6 \times 4$ $k$-point mesh. The results for Fe were used in the LSWT calculations of all configurations, while for Cu we used the value corrected by quantum fluctuation and ring-exchange ($J'^{eff}_{CuCu}=-8$ meV, see Supplementary Information~3).

\begin{table}[h!]
\centering
\renewcommand{\arraystretch}{1.5}
\begin{tabular}{| c || c | c |} 
 \hline
                  & $\qquad$  monoatomic planes $\qquad$   & diatomic checkerboard planes \\ [0.5ex] 
 \hline\hline
$ J^{ab}_{FeFe} (meV)$  & 8.76  &      {}  \\ 
 \hline
$ J^{ab}_{CuCu} (meV)$ & 130.62 &   {}  \\
 \hline
$J^{ab}_{CuFe} (meV)$    & {}       & 28.22   \\
 \hline
$J^{\perp}_{inter} (meV)$   & 1.39 &   1.32  \\
 \hline
$J^{\perp}_{intra} (meV)$   & -1.57 &  -1.59 \\
\hline
$J^{'}_{CuCu} (meV)$   & - &   24.53$^*$ \\
\hline
$J^{'}_{FeFe} (meV)$   & - &  0.65   \\ [1ex] 
 \hline
\end{tabular}
\caption{DFT+U exchange coupling parameters used in the LSWT calculations. $J>0$ indicates AFM coupling. *For $J^{'}_{CuCu}$, in LSWT we use the effective value $J'^{eff}_{CuCu}=-8$ meV (see Supplementary Information).  }
\label{TableJsMethods}
\end{table}

\paragraph*{LSWT and ASD dynamics calculations}

LSWT and ASD calculations are performed using the Julia package Sunny.jl~\cite{Dahlbom2025}. 
For LSWT calculations in Fig.~\ref{FigLSWT} the total $\mathscr{S}(\mathbf{q}, \omega)$ and projected $\mathscr{S}_{Cu}(\mathbf{q}, \omega)$, $\mathscr{S}_{Fe}(\mathbf{q}, \omega)$  are broadened by 50 meV and normalized by the total ($S_{tot}=3.0$),  the copper ($S_{Cu}=0.5$) and the iron spin quantum numbers ($S_{Fe}=2.5$), respectively. 
For ASD calculations, the time evolution of the spins  is calculated according to the stochastic Landau-Lifshitz equation:
\begin{equation}
\frac{\partial \mathbf{S}_i}{\partial t} = - \mathbf{S}_i \times \left[ \nabla E_i + \mathbf{\xi}_i(t) -  \frac{\alpha_{ii}}{{S}_i} \mathbf{S}_i \times  \nabla E_i \right]
\end{equation}
Here $\mathbf{B}_i=-1/(\gamma \hbar) \nabla E_i$ is the local field, $\mathbf{\xi}_i(t)$   is the fluctuating field. Thermalization is achieved by ensuring $\langle \xi_i(t) \xi_j(t') \rangle S_i= \alpha_{ii} k_B T \delta_{ij}\delta(t-t')$.
Integration is performed using the spherical implicit midpoint, with a step size of $dt=0.3$ fs. 
We consider supercells of $30\times30\times8$ atoms, and we average over $\approx100$ independent samples of the thermalized spin configuration.  
For calculations with $\langle dS \rangle \leq 1$  we start with a large super-cell (typically containing $\geq 5000$  Fe/Cu atoms) for the  ordered monoatomic configuration ($\langle dS \rangle=0$) and randomly substitute a given fraction of randomly chosen Cu (Fe) atoms in a given layer with Fe (Cu). For $1 \leq \langle  dS \rangle \leq 2 $, we start from the checkerboard ordered phase ($\langle dS \rangle=2$), pick  random Cu-Fe couples in the same plane and switch their position. In both cases the Cu--O--Fe bipyramids are conserved, so the corresponding atoms in the bipyramid are also modified. Couplings are tracked and updated both in- and out-of-plane. We then average over different ($\sim 20$) random realizations of the configuration for a given $\langle dS \rangle$.
The calculated dynamical time correlation function $\mathscr{S}(\mathbf{r}, t)$ is then Fourier transformed to obtain $\mathscr{S}(\mathbf{q}, \omega)$. For the calculations of Fig.~\ref{FigFWHM}, we used DHO fitting to extract frequencies, intensities and peak broadening $2\gamma$.
Simulations for magnetic and lattice Bragg reflection intensities are obtained by calculating the magnetic structure in the entire supercell and then averaging over $\sim20$ random configurations.

\paragraph*{Entropy calculations}
We consider each bipyramid double-layer with $N$ total Fe and Cu atoms. First, we allow random in-plane occupation but without Fe--O--Fe and Cu--O--Cu impurities.
The configurational entropy $S_{conf}$ as a function of $\langle dS \rangle$ is calculated as:

\begin{align}
    S_{conf}&= k_B \ln(\frac{M!}{xM!(1-x)M}) \approx \frac{N}{2}k_B \ln\left[ + (x)\ln(x) + (1-x)\ln(1-x)\right]. \nonumber
\end{align}
where $M=N/2$ the number of atoms per layer.

For $\langle dS \rangle \leq 1$,  $x$ denotes the fraction of randomly placed Fe atoms in the bottom Cu layer. With this definition it is straightforward to get the relation  $x={\langle dS \rangle}/{2}$.

For the case $1 < \langle dS \rangle \leq 2$, $x$ indicates the fraction of Fe atoms in the bottom layer that, starting from a checkerboard order,  have been  randomly swapped with  Cu atoms in the same layer. For this case, it can be shown that $x=(1  -  \sqrt{\langle dS \rangle -1})/2$.

Then, we consider the inclusion of $yM$ Fe--O--Fe (and Cu--O--Cu) impurities for the case $\langle dS \rangle \leq 1$ , for which we get:

\begin{align}
    \frac{S_{conf}+S_{imp}}{k_B} = &\ln (\frac{M!}{m_{FeFe}! \times m_{FeCu}! \times m_{CuFe}! \times m_{CuCu}!}) \nonumber \\
    \approx& - \frac{N}{2} [2y \ln(y) + (x-y)\ln(x-y)    + (1-x-y)\ln(1-x-y)] \nonumber
\end{align}
where, for instance, $m_{FeCu}$ is the number of bipyramids with Fe atom at the bottom and Cu at the top.
With this convention, the $T_{spiral}$ of our crystal corresponds to $y=0.06$~\cite{Scaramucci2020} resulting in ${(S_{conf}+S_{imp})}/{N k_B}\approx 0.53$  when impurities are taken into account in the $x=0.5$ mixed phase.

\section*{X-ray diffraction and elastic neutron scattering results} \label{CameaXrayDiffraction}

Powder x-ray diffraction data were collected on a Bruker D8 Advance diffractometer using Cu $K\alpha$ radiation at room temperature. A finely crushed fragment of the single crystal was spread on a flat aluminum sample holder for measurement. The results are shown in Extended Data Fig.~\ref{XrayDiffrCamea}\textbf{a}, together with calculated patterns obtained using the FullProf suite~\cite{RODRIGUEZCARVAJAL199355} for the monoatomic (configuration in Fig.\ref{FigLSWT}~\textbf{a}) and diatomic checkerboard (configuration in Fig.\ref{FigLSWT}~\textbf{e}) configurations. No superstructure Bragg reflections, such as the $(0.5,0.5,1.0)$ peak expected at $Q\sim 1.41$~\AA$^{-1}$ for the checkerboard configuration, are observed in the experimental pattern.

Elastic neutron scattering data, collected at the CAMEA neutron spectrometer~\cite{Camea2023} at $T=200$~K, are shown in Extended Data Fig.~\ref{XrayDiffrCamea}\textbf{b}. The scattering intensity along the high symmetry line $[H,H,1.5]$ r.l.u. shows a single peak at the CM wavevector $(0.5, 0.5, 1.5)$, with no evidence of the additional peak at $(0,0,1.5)$ predicted by ASD calculations for the diatomic checkerboard configuration.

\clearpage
\section*{Extended Data}

\renewcommand{\figurename}{Extended Data Fig.}
\setcounter{figure}{0}

\begin{figure}[h]
    \centering
    \includegraphics[width=\textwidth]{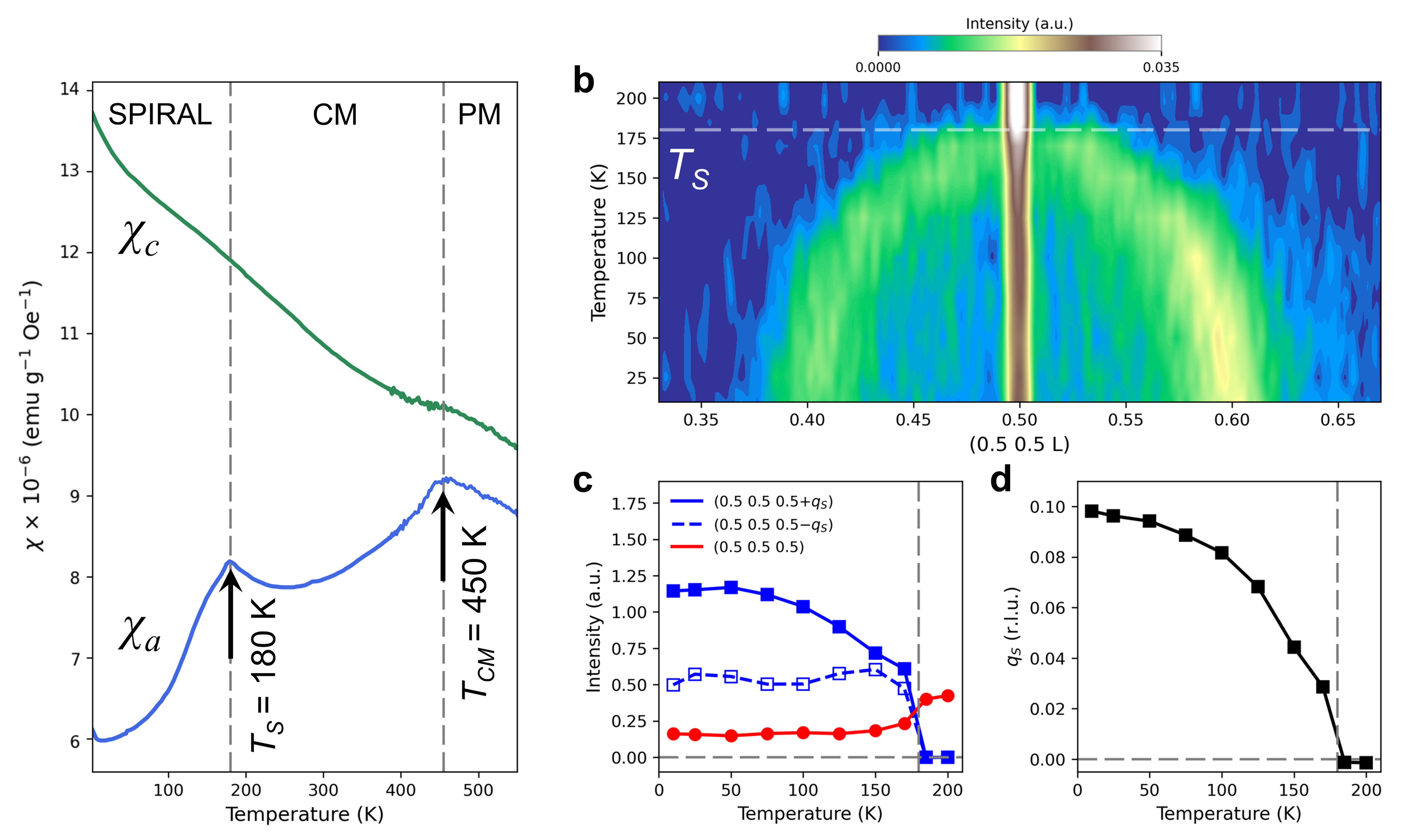}
    \caption{\textbf{Magnetic susceptibility and neutron diffraction characterization of the YBCFO single crystal.}
        \textbf{a}, DC magnetic susceptibility $\chi=M/H$ measured under field-cooled (FC) conditions with $\mu_0H=0.2$~T applied along the crystallographic $a$ and $c$ axes. The anomalies at $T_{CM}=450$~K and $T_S=180$~K mark the transitions to the commensurate antiferromagnetic (CM) and incommensurate (ICM) spiral phases, respectively.
        \textbf{b}, Temperature-dependent neutron diffraction intensity map measured along $\mathbf{Q}=(0.5,\,0.5,\,L)$ in the vicinity of $(1/2,1/2,1/2)$ magnetic Bragg reflection.
        \textbf{c}, Integrated neutron diffraction intensities of the commensurate peak at $\mathbf{Q}=(0.5,0.5,0.5)$ and of the incommensurate satellite peaks at $\mathbf{Q}=(0.5,0.5,0.5\pm q_S)$ as a function of temperature.
        \textbf{d}, Temperature dependence of the incommensurability $q_S$ extracted from pseudo-Voigt fits to the satellite peak positions.}
    \label{Fig_Transitions}
\end{figure}

\begin{figure}[h]
    \centering
    \includegraphics[width=\textwidth]{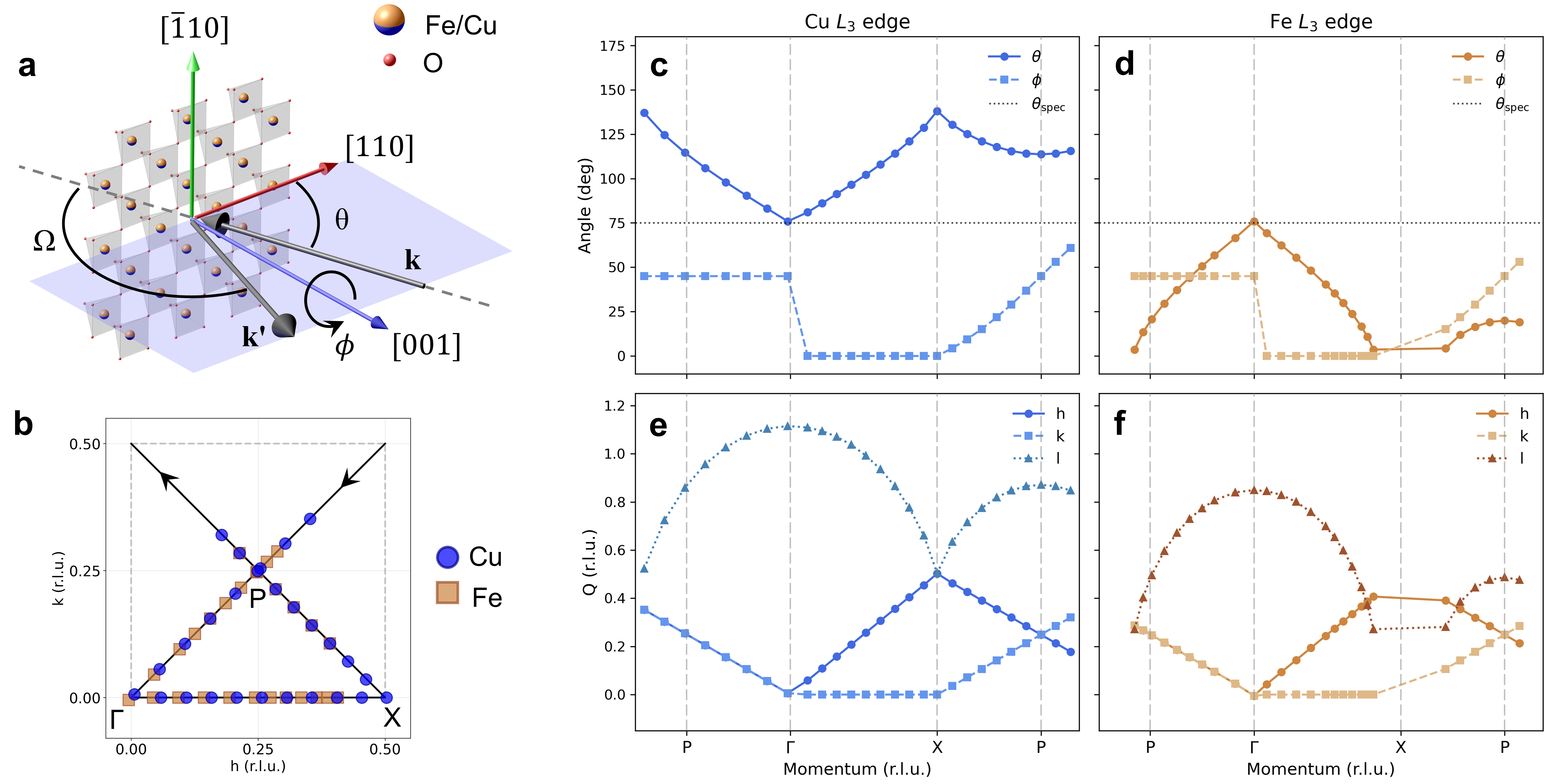}
    \caption{\textbf{RIXS geometry and reciprocal-space trajectories.}
    \textbf{a}, Experimental scattering geometry for the [001]-oriented YBCFO crystal. The sample orientation is controlled through the incidence angle $\theta$ and azimuthal angle $\phi$ at fixed scattering angle $\Omega$.
    \textbf{b}, Measured momentum-space path in the projected $(h,k)$ plane, connecting the high-symmetry points $P$, $\Gamma$, $X$, and $P$. The Cu/Fe checkerboard motif of the rotated $\sqrt{2}\times\sqrt{2}\times 2$ supercell used in the LSWT calculations is overlaid for reference.
    \textbf{c,d}, Sample angles $\theta$ and $\phi$ at the Cu $L_3$ and Fe $L_3$ edges.
    \textbf{e,f}, Corresponding momentum coordinates $(h,k,l)$ along the experimental path. Vertical dashed lines mark the high-symmetry points.}
    \label{Fig_RIXS_Geometry}
\end{figure}

\begin{figure}[h]
    \centering
    \includegraphics[width=\textwidth]{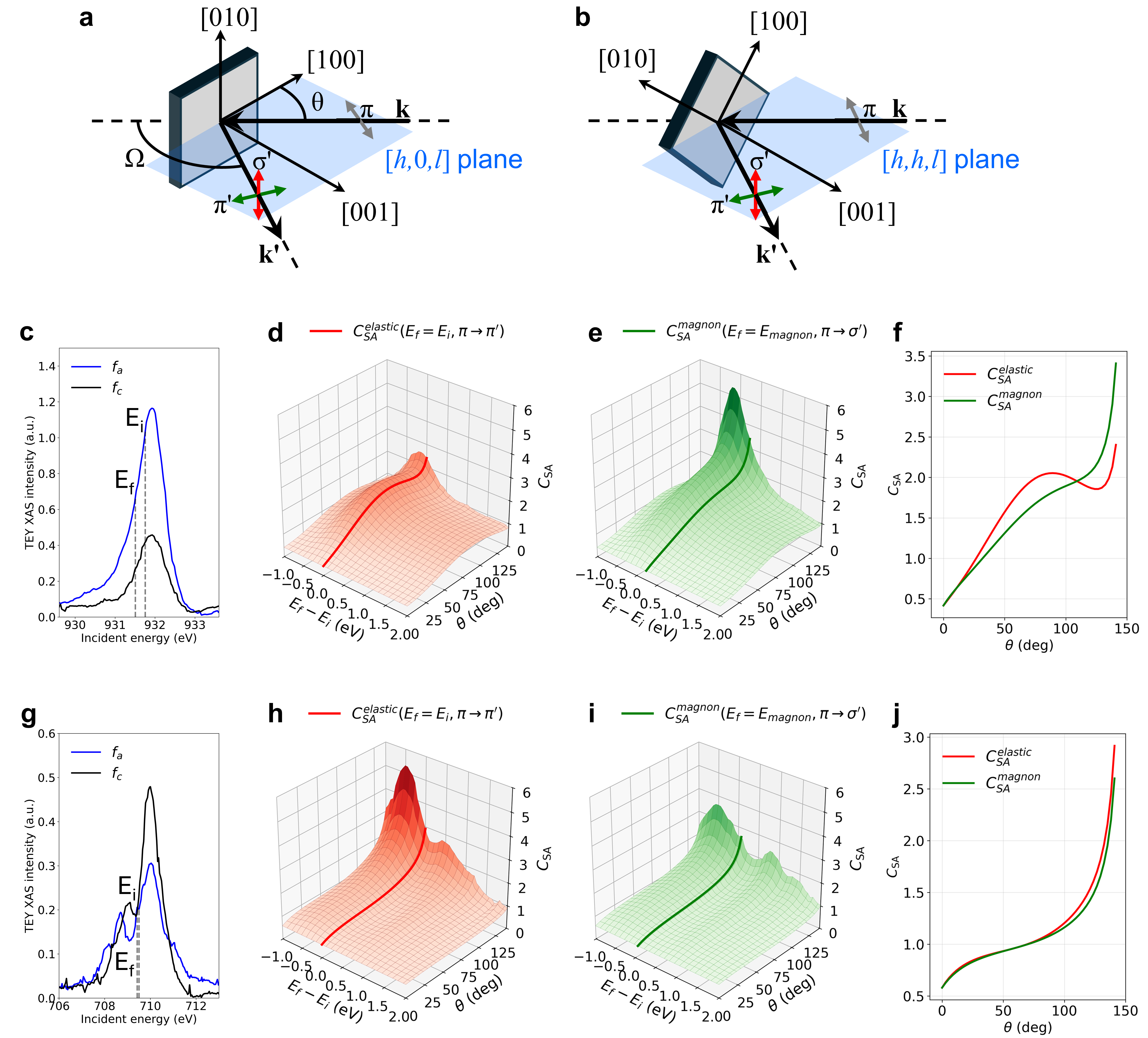}
    \caption{\textbf{XAS-derived self-absorption correction for Cu and Fe $L_3$-edge RIXS.}
    \textbf{a,b}, Scattering geometries used for RIXS measurements in the $[h,0,l]$ and $[h,h,l]$ planes, respectively. The incident $\pi$ polarization, scattered wavevector $\mathbf{k}'$, crystallographic directions, and the polarization directions of scattered photons used for the derivation of self-absorption coefficients are indicated.
    \textbf{c}, Orientation-dependent TEY XAS spectra measured at the Cu $L_3$ edge with the electric field predominantly aligned along the crystallographic $a$ and $c$ directions, yielding the absorption form factors $f_a(E)$ and $f_c(E)$. The dashed lines mark the incident photon energy ($E_i$) and the overall scattered photon energies for magnon peaks ($E_f$) in the RIXS measurements.
    \textbf{d,e}, Calculated self-absorption factor $C_{\mathrm{SA}}$ at the Cu $L_3$ edge for the elastic/phonon ($\pi\rightarrow\pi'$) and magnetic ($\pi\rightarrow\sigma'$) channels, shown as a function of energy loss and incidence angle $\theta$.
    \textbf{f}, Angular cuts of the Cu $L_3$ self-absorption factor at the selected emitted energies for the elastic/phonon and magnetic channels.
    \textbf{g}, Orientation-dependent TEY XAS spectra measured at the Fe $L_3$ edge, yielding the absorption form factors $f_a(E)$ and $f_c(E)$. The dashed lines illustrate the incident photon energies ($E_i$) and overall scattered photon energies for magnon peaks ($E_f$) in the RIXS measurements.
    \textbf{h,i}, Calculated self-absorption factor $C_{\mathrm{SA}}$ at the Fe $L_3$ edge for the elastic/phonon ($\pi\rightarrow\pi'$) and magnetic ($\pi\rightarrow\sigma'$) channels, shown as a function of energy loss and incidence angle $\theta$.
    \textbf{j}, Angular cuts of the Fe $L_3$ self-absorption factor at the selected 
    emitted energies for the elastic/phonon and magnetic channels.}
    \label{Fig_XAS_SA}
\end{figure}

\begin{figure}[h]
    \centering
    \includegraphics[width=\textwidth]{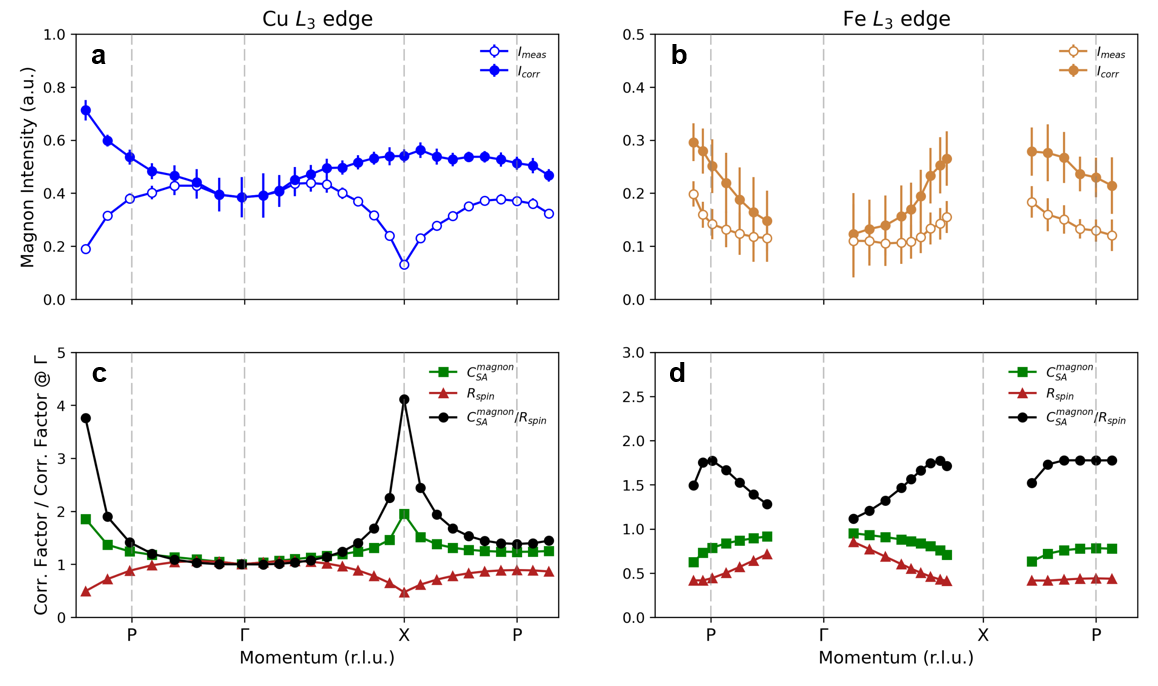}
    \caption{\textbf{Correction of the integrated magnon intensities.}
    \textbf{a,b}, Integrated magnon intensities extracted from the DHO fits before ($I_{\mathrm{meas}}$) and after ($I_{\mathrm{corr}}$) applying the combined self-absorption and spin-flip cross-section correction, for the Cu $L_3$ and Fe $L_3$ edges, respectively.
    \textbf{c,d}, Calculated magnetic self-absorption factor $C_{\mathrm{SA}}^{\mathrm{mag}}$, spin-flip factor $R_{\mathrm{spin}}$, and total correction factor $C_{\mathrm{SA}}^{\mathrm{mag}}/R_{\mathrm{spin}}$ along the measured momentum trajectory for the Cu $L_3$ and Fe $L_3$ edges, respectively. All correction factors are normalized to unity at $\Gamma$. Vertical dashed lines mark the high-symmetry points of the measured path.}
    \label{Fig_IntensityCorrections}
\end{figure}

\begin{figure}[h]
    \centering
    \includegraphics[width=\textwidth]{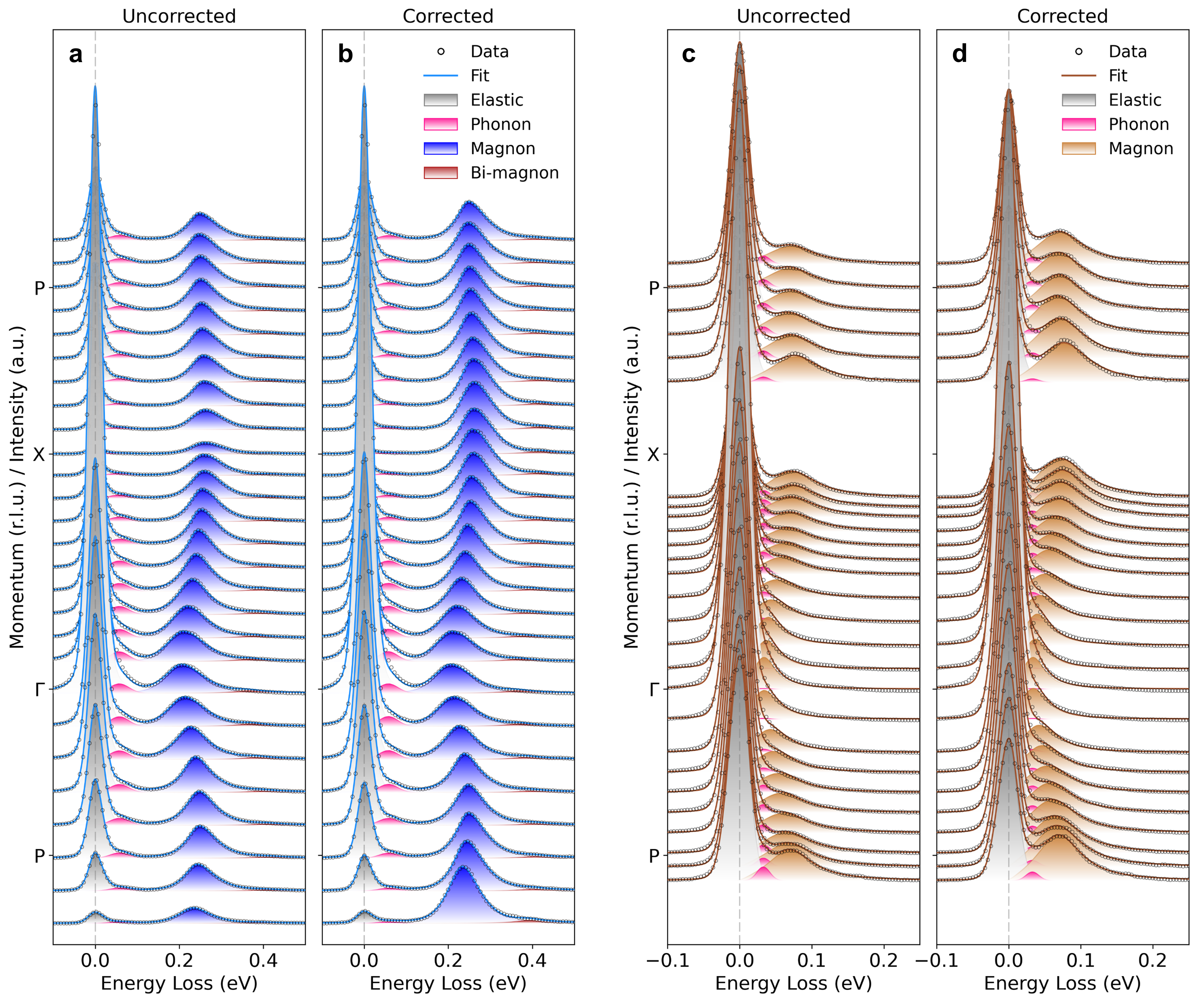}
    \caption{\textbf{Representative uncorrected and corrected RIXS spectral stacks.}
    \textbf{a,b}, Cu $L_3$-edge spectra plotted along the measured momentum trajectory before and after applying the energy-dependent effective correction factor $C_\mathrm{eff}(\omega)$, respectively. The fitted contributions from the elastic line, phonon, magnon, and bimagnon channels are indicated.
    \textbf{c,d}, Fe $L_3$-edge spectra plotted before and after applying the same correction procedure. The fitted elastic, phonon, and magnon contributions are indicated. The corrected spectra are used for the visualization of the momentum-energy maps shown in Fig.~\ref{FigRIXS}.}
    \label{Fig_Corrected-Spectra}
\end{figure}

\begin{figure}[t]
    \centering
    \includegraphics[width=0.5\columnwidth]{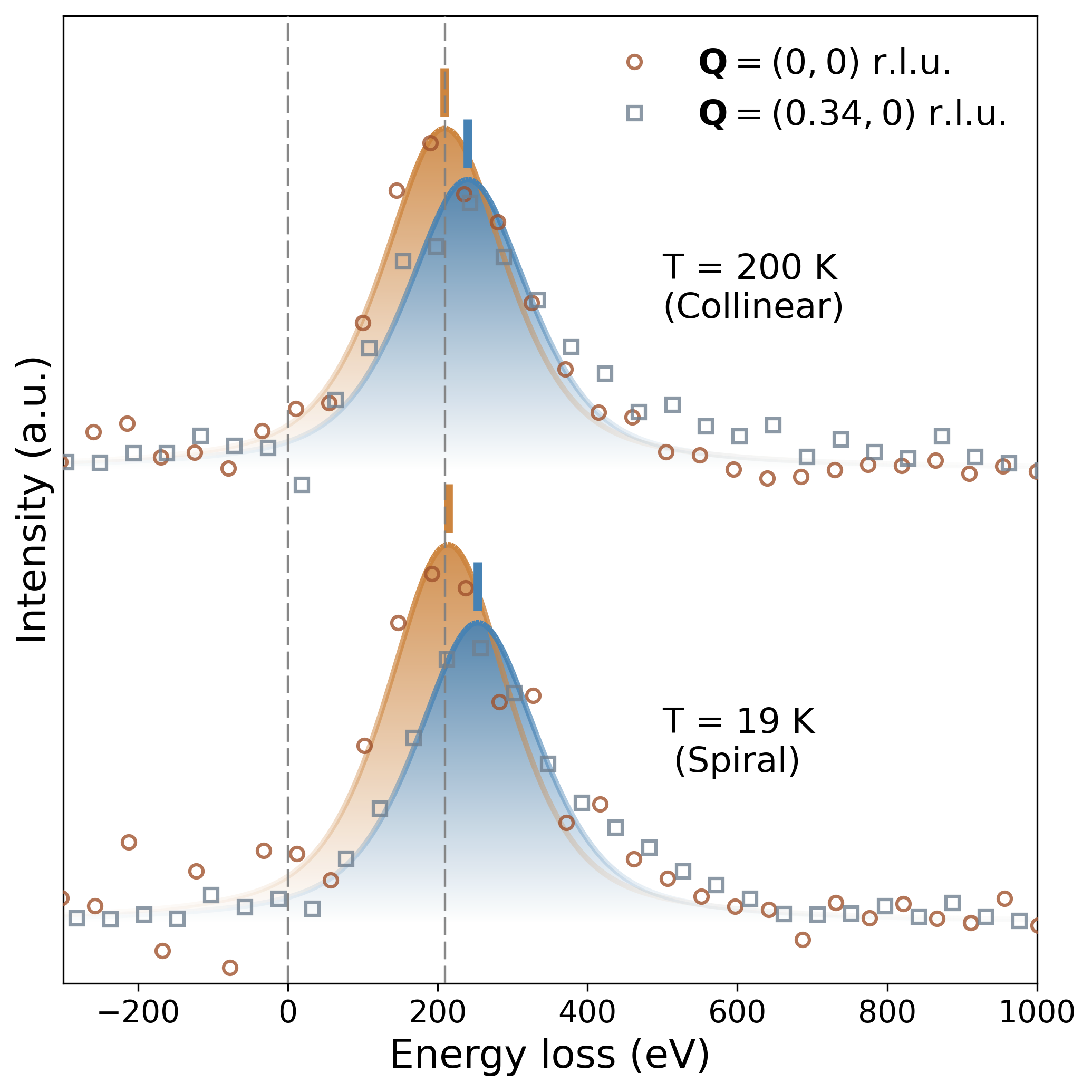}
    \caption{\textbf{Temperature dependence of the optical magnon branch measured at the Cu $L_3$ edge.} RIXS spectra collected at the Furka endstation of SwissFEL at $\mathbf{Q}=(0,0)$ (brown circles) and $\mathbf{Q}=(0.34,0)$~r.l.u. (gray squares) in the collinear commensurate phase ($T=200$~K, top) and the spiral phase ($T=19$~K, bottom). Solid lines are pseudo-Voigt fits to the data. The elastic line contribution has been subtracted for clarity. Vertical dashed lines mark zero energy loss and the fitted peak position at $\Gamma$. The magnon energy and lineshape show no significant change across the commensurate-to-spiral transition at $T_S=180$~K, consistent with the preserved in-plane spin order in both phases.}
    \label{fig_Tdep_Furka}
\end{figure}

\begin{figure}[t]
    \centering
    \includegraphics[width=\columnwidth]{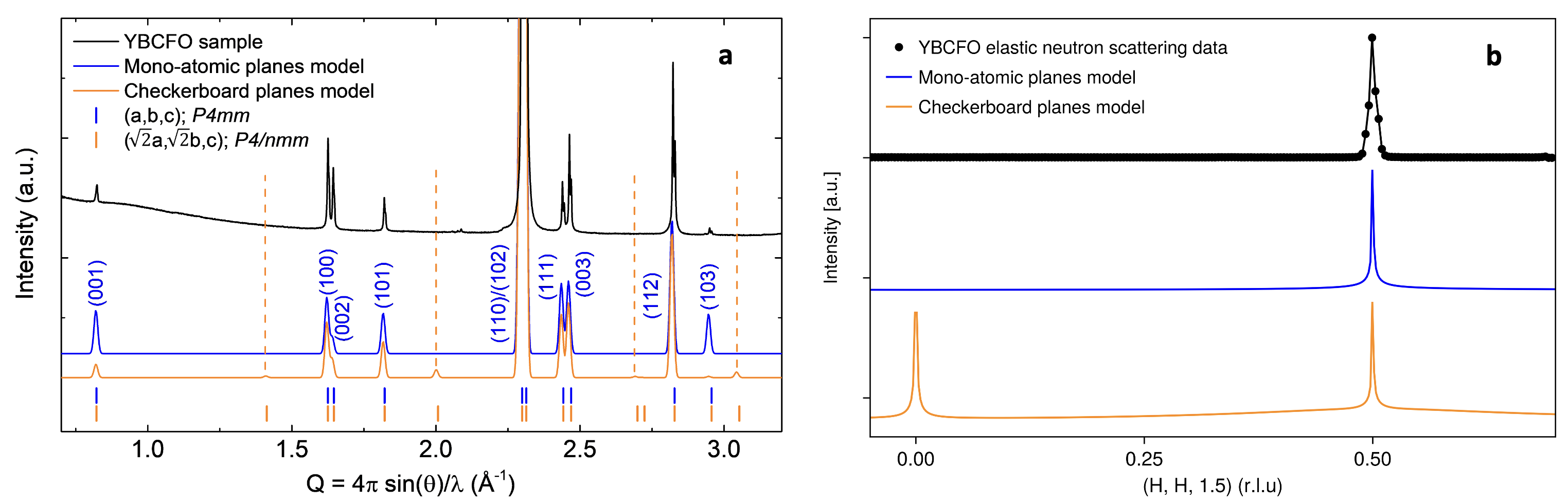}
    \caption{\textbf{X-ray diffraction and elastic neutron scattering characterization.} \textbf{a}, Measured x-ray diffraction intensity (black) compared with calculated structure factors for two representative Fe/Cu arrangements: mono-atomic Fe/Cu planes (blue) and checkerboard-ordered Fe/Cu planes (orange). The experimental pattern shows no additional superstructure Bragg reflections (dashed vertical lines), ruling out long-range in-plane bi-atomic Fe/Cu ordering within the $ab$ planes. The vertical line markers indicate the reflection positions calculated for the two idealized structural models: blue corresponds to a $P4mm$ cell indexed in the parent tetragonal basis $(a,b,c)$ (Fig.~\ref{FigLSWT}\textbf{a}), whereas orange corresponds to a checkerboard-ordered $P4/nmm$ superstructure indexed in the rotated $(\sqrt{2}a,\sqrt{2}b,c)$ unit cell (Fig.~\ref{FigLSWT}\textbf{e}). \textbf{b}, Elastic neutron scattering data (top), and simulations for mono-atomic planes (middle) and checkerboard planes (bottom).}
    \label{XrayDiffrCamea}
\end{figure}

\clearpage

% ---- Supplementary Material ----
\newpage

\newgeometry{left=3.0cm, right=3.0cm, top=2.5cm, bottom=2.5cm}

\linespread{1.5}
\frenchspacing

%% Supplementary numbering
\renewcommand{\thefigure}{S\arabic{figure}}
\renewcommand{\thetable}{S\arabic{table}}
\renewcommand{\theequation}{S\arabic{equation}}
\setcounter{figure}{0}
\setcounter{table}{0}
\setcounter{equation}{0}

\renewcommand{\vec}[1]{\mathbf{#1}}
\newcommand{\gvec}[1]{\boldsymbol{#1}}

\begin{center}
{\Large \textbf{Supplementary Information}}\\[6pt]
{\large Coherent spin waves in a maximal entropy phase}\\[6pt]
Arnau~Romaguera$^{x}$, Eugenio Paris, Elizabeth Skoropata,
Stefano Agrestini, Mirian Garcia-Fernandez, Marisa Medarde,
Noah Schnitzer, Lopa Bhatt, Berit H. Goodge, Yun Yen,
Matthias Krack, Michael Schüler, Romain Sibille,
Tom Fennell, Daniel G. Mazzone, Jakob Lass, Ellen Fogh,
Anirudha Ghosh, Marco Caputo, Carlos William Galdino, Zhijia Zhang,
Thorsten Schmitt, Milan Radovic, Luc Patthey, Hiroki Ueda,
Monica Ciomaga Hatnean,  Elia~Razzoli$^{*}$\\[4pt]
\small $^{*}$Correspondence: elia.razzoli@psi.ch
\small $^{x}$Correspondence: arnau.romaguera@psi.ch
\end{center}

\bigskip
\noindent\textbf{This file includes:}\\
Supplementary Information 1--5\\
Supplementary Table~\ref{TableJsSuppl}\\
Supplementary Figs.~\ref{FigS1}--\ref{FigSpiral}\\

\newpage

\section{LSWT in YBCFO: Numerical results}

\begin{figure*}[t]
    \centering
    \includegraphics[width=\columnwidth]{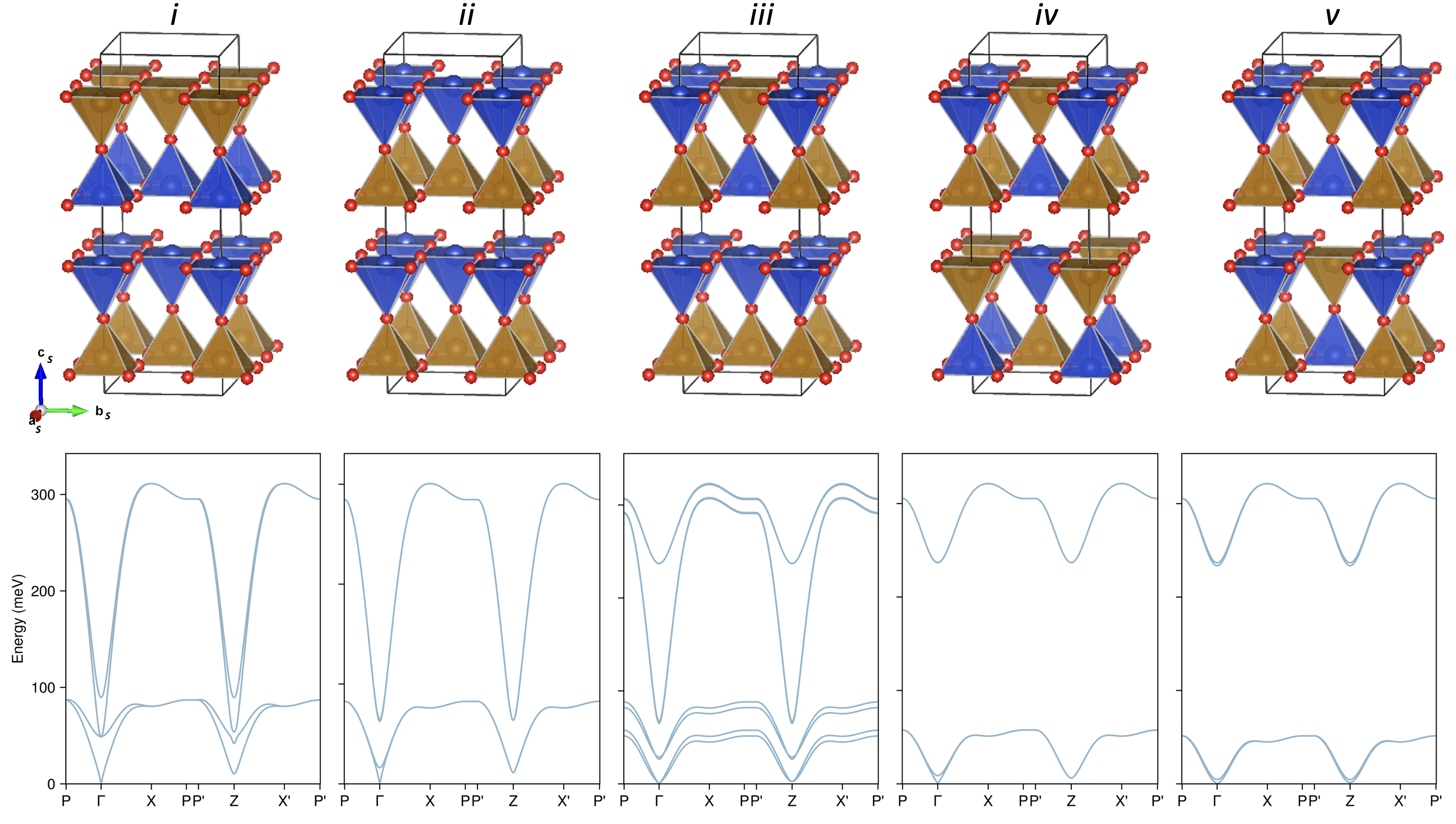}
    \caption[FigS1]{Linear Spin wave theory calculations for various configurations of atomic ordering in YBCFO. Red, brown and blue are Oxygen, Iron and Copper atoms, respectively. (i)-(v) Low energy atomic configurations (from~\cite{Morin2015_S}). Drawings generated using VESTA software~\cite{momma2011vesta_S}. (f)-(g) Spin wave dispersion along high symmetry lines $\Gamma = (0, 0, 0)$, $\text{P}=(0.25,0.25,0)$,  $\text{X}=(0.5,0,0)$, $Z {'} = (0, 0, 0.5)$, $\text{P}{'}=(0.25,0.25,0.5)$ and $\text{X'}=(0.5,0,0.5)$  in r.l.u [units of ($2\pi/a$, $2\pi/b$, $2\pi/c$)].}
    \label{FigS1}
\end{figure*}

In Table~\ref{TableJsSuppl} we summarized the DFT+U calculated couplings for all the low energy configurations in YBCFO. When multiple couplings are present for the same bond, due to a different atomic environment, both values are given. 
\begin{table}[h!]
\centering
\begin{tabular}{| c || c | c | c | c | c |} 
 \hline
                  &  i   & ii  & iii        & iv   & v \\ [0.5ex] 
 \hline\hline
$ J^{ab}_{FeFe} (meV)$ & 8.62    & 8.76  &   8.7      &    {}  &    {}  \\ 
 \hline
$ J^{ab}_{CuCu} (meV)$ & 135.02  & 130.62 & 133.67       &    {}  &    {}  \\
 \hline
$J^{ab}_{CuFe} (meV)$ & {}     & {}       & 28.41, 28.1   &   28.16 & 28.22   \\
 \hline
$J^{c}_{inter, FeFe} (meV)$ & 2.68     & {} & 2.89          &   3.01  &    {}  \\
 \hline
$J^{c}_{inter, CuCu} (meV)$ & 10.63   & {} & 8.99           &   7.58  &    {}  \\
 \hline
$J^{c}_{inter, CuFe} (meV)$ & {}     & 1.39 & 1.07, 1.69     &   {}   &   1.32  \\
 \hline
$J^{c}_{intra} (meV)$  & -1.58  & -1.57 & -1.7, -1.47 &  -1.61 &  -1.59 \\
\hline
$J^{'}_{CuCu} (meV)$  & -  & - & - &  - &  24.53$^*$ \\
\hline
$J^{'}_{FeFe} (meV)$  & - & - & - &  - &  0.65   \\ [1ex] 
 \hline
\end{tabular}
\caption{Exchange coupling parameters calculated in DFT+U and used in the LSWT calculations. $J>0$ indicates AFM coupling. *For $J^{'}_{CuCu}$, in LSWT we use the effective value $J'^{eff}_{CuCu}=-8$ meV (see section \ref{ring}).  }
\label{TableJsSuppl}
\end{table}
To describe the idealized low-energy Cu/Fe atomic configurations shown in Fig.~\ref{FigS1}, we adopt a rotated $\sqrt{2}\times\sqrt{2}\times 2$ supercell whose basis vectors are related to those of the parent tetragonal cell by
$(\mathbf a_s,\mathbf b_s,\mathbf c_s)=(\mathbf a,\mathbf b,\mathbf c)\,P$ with transformation matrix
\begin{equation}
P=
\begin{pmatrix}
1 & -1 & 0\\
1 & \phantom{-}1 & 0\\
0 & \phantom{-}0 & 2
\end{pmatrix}.
\end{equation}
That is: $\mathbf{a}_s=\mathbf{a}+\mathbf{b}$, $\mathbf{b}_s=-\mathbf{a}+\mathbf{b}$ and $\mathbf{c}_s = 2\mathbf{c}$. Accordingly, $a_s=b_s=\sqrt{2}\,a$ and $c_s=2c$.  
The configurations we used in the main text are \textit{ii} and \textit{v}.  Calculations performed with Sunny.jl~\cite{Dahlbom2025_S} are shown at the bottom of Fig.~\ref{FigS1}.
 In configurations \textit{i}-\textit{ii} each 2D-plane is composed by a single atomic type (either Fe or Cu), with the main difference given by the stacking order between transition metal layers. In this case, we expect for each layer to follow the characteristic dispersion of magnons in monoatomic square lattice $\omega(\mathbf{q}) =4JS \sqrt{1- \frac{1}{4}[\cos(q_x)+\cos(q_y)]^2}$~\cite{RevModPhys.63.1}, with $4JS\approx260$ meV and $4JS\approx90$ meV, for Cu and Fe layers, respectively. Degeneracy at $\Gamma$ is lifted by interlayer couplings, giving rise to a small gap ($<100$ meV) between optical and acoustic branches.
In \textit{iv}-\textit{v} each 2D-plane contains both Cu and Fe atoms, and the magnon dispersion is the one of weakly coupled diatomic square lattice planes, and thus acoustic and optical branches are expected to generate in the 2D layers dispersion, i.e., already in the absence of inter-layer couplings. 
Finally, in \textit{iii}, both monoatomic and diatomic planes are present, and the magnon dispersion presents the main features of both \textit{i}-\textit{ii} and  \textit{iv}-\textit{v} configurations.

\section{LSWT in YBCFO: Analytical results }
\subsection{ Optical and acoustic branch in the 2D limit }

We analyze first the simple case of a single 2D plane with general $S_A$, $S_B$ on each sublattices. The AFM nearest-neighbors and FM next-nearest-neighbors are indicated as $J$ and $J'$. To obtain the magnon dispersion we use the Holstein-Primakoff transformations with $S_A$ and $S_B$ (to first order in 1/S): 
\begin{align*}
[S^-_{Ai}, S^+_{Ai}, S^z_{Ai}]&= [\sqrt {2S_A} a_i^\dag, \sqrt {2S_A} a_i, S_A - a_i^\dag a_i ] \\
[S^-_{Bj}, S^+_{Bj}, S^z_{Bj}]&= [\sqrt {2S_B} b_j, \sqrt {2S_B} b_j^\dag, -S_B + b_j^\dag b_j ] 
\end{align*}
with $z-$direction aligned along the spin-ordering direction $a$.
In $\mathbf{q}$ space, the Hamiltonian becomes:
\begin{align*}
H &= const \\
&+ Jz \sum_\mathbf{q} (S_B + S_A \rho + H_0) a_\mathbf{q}^\dag a_\mathbf{q} \\
&+   (S_A+S_B \rho - H_0) b_\mathbf{q}^\dag b_\mathbf{q} \\
&+ \sqrt{S_AS_B} \gamma^{\parallel}_{\mathbf{q}} (a_\mathbf{q} b_{-\mathbf{q}} + a_\mathbf{q}^\dag b_{-\mathbf{q}}^\dag)
\end{align*}
where $z$ is the coordination number, $\rho= - J'(1- \gamma{'}_{\mathbf{q}})/Jz$ and $\gamma^{\parallel}_\mathbf{q}=\frac{1}{2} [\cos(2\pi q_x)+\cos(2\pi q_y)]$ and $\gamma_\mathbf{q}'=\cos(2\pi q_x)\cos(2\pi q_y)$.
We then employ a Bogoliubov transformation, to keep the bosonic commutation rules and to diagonalize the Hamiltonian:
\begin{align*}
\begin{pmatrix}
\alpha_\mathbf{q}  \\
\beta^\dag_{-\mathbf{q}}
\end{pmatrix}
&=
\begin{pmatrix}
+u_\mathbf{q} & - v_\mathbf{q}  \\
-v_\mathbf{q} & + u_\mathbf{q} 
\end{pmatrix}
\begin{pmatrix}
a_\mathbf{q}  \\
b^\dag_{-\mathbf{q}}
\end{pmatrix} 
= \mathbf{U}\begin{pmatrix}
a_\mathbf{q}  \\
b^\dag_{-\mathbf{q}}
\end{pmatrix} 
\end{align*}
with $u_\mathbf{q}^2 - v_\mathbf{q}^2 = 1$.
Diagonalization is equivalent to ask for $[\alpha_q, H] = \omega_q \alpha_q$ and $[\beta_q, H] = \omega_q \beta_q$ (and c.c. equations) to obtain the optical and acoustic branch respectively, so we have for the equation for $\alpha_\mathbf{q}$:
\begin{align*}
\begin{pmatrix}
S_B+ S_A \rho + H_0 -\frac{\omega_q}{Jz}  & \sqrt {S_A S_B} \gamma_q  \\
\sqrt {S_A S_B} \gamma_q & S_A+S_B \rho - H_0 + \frac{\omega_q}{Jz}
\end{pmatrix}
\begin{pmatrix}
u_\mathbf{q} \\
v_\mathbf{q} 
\end{pmatrix} 
&= \\
(\mathbf{H} - \frac{\omega_q}{Jz} \mathbf{g}) 
\begin{pmatrix}
u_\mathbf{q} \\
v_\mathbf{q} 
\end{pmatrix}
&=0 
\end{align*}
which has non trivial solution for  $\det (\mathbf{H} - \frac{\omega_q}{Jz} \mathbf{g}) \neq 0$ i.e., (taking the positive root, for $S_B>S_A$):
\begin{align}
\frac{\omega^{opt}(\mathbf{q})}{Jz} = &H_0 + dS_K/2 + \omega_0
\label{SEQ1}
\end{align}
with
\begin{align*}
dS_K &= (S_B - S_A)(1- \rho) = dS(1-\rho)  \\ 
S_{avg} &= \frac{S_A+S_B}{2}, \quad S_{geo} = \sqrt{S_AS_B}  \\  
S_{rms} &= \sqrt{\frac{(S_A)^2+(S_B)^2}{2}}  \\
\omega_0 &= \frac{1}{2}\sqrt{dS_K^2 + 4 S^2_{geo}( 1- \gamma_q^2 + \rho^2) + 8S_{rms}^2 \rho }\\
&= \sqrt{S_{avg}^2(1+\rho)^2 -  S_{geo}^2 \gamma_q^2  }
\end{align*}

Similarly for the equation for $\beta_q$ we obtain:
\begin{align}
\frac{\omega^{ac}(\mathbf{q})}{Jz} = &-(H_0 + dS_K/2) + \omega_0
\label{SEQ2}
\end{align}
As expected eq. (\ref{SEQ1})-(\ref{SEQ2}) reduce to the well-known equation for AFM spin waves for $dS=0$, which implies that geometrical,  root mean square (rms) averages are all equal to $S_A=S_B=S$~\cite{Rezende2019}. It also illustrates the similar effect of external field $B_0$ and $dS$ in opening a gap in the magnon spectrum. 
\\
The values of $u_\mathbf{q}$, and $v_\mathbf{q}$ are obtained by setting the off diagonal elements of $\mathbf{U}^T\mathbf{H}\mathbf{U}$ to zero and are given by:
\begin{align}
u_\mathbf{q} &= \sqrt{\frac{1}{2} + \frac{S_{avg}(1-\rho)}{2 \omega_0} }  \\
v_\mathbf{q} &= -\text{sgn}(\gamma_\mathbf{q})\sqrt{-\frac{1}{2} + \frac{S_{avg}(1-\rho)}{2 \omega_0} }   
\label{SEQ9}
\end{align}

To calculate $\mathscr{S}(\mathbf{q}, \omega)= \mathscr{S}^{xx}(\mathbf{q},\omega)+\mathscr{S}^{yy}(\mathbf{q},\omega)+\mathscr{S}^{zz}(\mathbf{q},\omega)$, with the definition with  $\mathscr{S}^{XX}=\langle (S^X_{A\mathbf{q}}+S^X_{B\mathbf{q}})(S^X_{A\mathbf{q}}+S^X_{B\mathbf{q}})\rangle$, and given that $\mathscr{S}^{XX}=\mathscr{S}^{YY}$, $\mathscr{S}^{ZZ}=0$  we obtain:
\begin{align*}
\mathscr{S}(\mathbf{q}, \omega) \propto &(S_Au_\mathbf{q}^2 + 2 \sqrt{S_AS_B} u_\mathbf{q}v_\mathbf{q} + S_Bv_\mathbf{q}^2) \delta(\omega - \omega^{opt}(\mathbf{q}))  \nonumber \\
  + &(S_Bu_\mathbf{q}^2 + 2 \sqrt{S_AS_B} u_\mathbf{q}v_\mathbf{q} + S_Av_\mathbf{q}^2) \delta(\omega - \omega^{ac}(\mathbf{q}))  
\end{align*}

At $(0,0)$, both acoustic and optical branches have an intensity minimum, with the latter having a vanishing intensity, i.e.,:
\begin{align}
R_I=\frac{I^{opt}(0, 0)}{I^{opt}(0.25,0.25)}&=0  
\end{align}
The intensity at $(0.5, 0.5)$, is maximal but finite.  

Projection on $A$ atom is achieved by calculating $S_A^{XX}= \langle S^X_{A\mathbf{q}}S^X_{A\mathbf{q}} \rangle$, which gives, for the optical (acoustic) branch $I^{opt}_{A}\propto S_A u_q^2$ ($I^{ac}_{A}\propto S_A v_q^2$). In the projected case,  both branches have the global intensity maxima at $(0,0)$ and $(0.5,0.5)$, and the intensity ratio is given by:
\begin{align}
R_I=\frac{I_{A}^{opt}(0, 0)}{I_{A}^{opt}(0.25,0.25)}&=\frac{u_{q=(0, 0)}^2}{u_{q=(0.25, 0.25)}^2}  \nonumber \\ 
&=\frac{S_{B}}{2(S_{B}-S_{A})}(1+\frac{1-\rho}{1+\rho}) \nonumber \\ 
&\approx \frac{S_B}{dS}
\end{align}
where we used $J' \ll J $ in the last line. For the case of A=Cu and B=Fe atoms we would then get $R_I\approx 1.25$.
\subsection{Optical and acoustic branch in 3D} \label{DispDerivation}

In this section, we discuss the three-dimension case of dispersion of the magnon branches in YBCFO. For simplicity, we consider the case of a $\sqrt{2}\times\sqrt{2}\times 1$ supercell of size $(a_s,b_s,c_s/2)$, which can be applied to configurations \textit{ii} and \textit{v} in Fig.~\ref{FigS1}.
In such a case, we consider four in-equivalent magnetic atoms in the magnetic unit cell with coordinates:
\begin{align*}
\mathbf{r_A}&= (0, 0, 0.25) \\
\mathbf{r_B}&= (0.5, 0.5, 0.25) \\
\mathbf{r_C}&= (0.5, 0.5, 0.75) \\
\mathbf{r_D}&= (0, 0, 0.75) \\
\end{align*}
We take the Hamiltonian for a Heisenberg model on a 3D lattice:
\begin{align*}
H = &J^{\parallel}_{AB} \sum_{i, j} \mathbf{S}_{Ai} \cdot \mathbf{S}_{Bj} +J^{\parallel}_{CD} \sum_{i, j} \mathbf{S}_{Ci} \cdot \mathbf{S}_{Dj} \\
&J^{'\parallel}_{AA} \sum_{<i, j>} \mathbf{S}_{Ai} \cdot \mathbf{S}_{Aj} +J^{'\parallel}_{BB} \sum_{<i, j>} \mathbf{S}_{Bi} \cdot \mathbf{S}_{Bj} \\
&J^{'\parallel}_{CC} \sum_{<i, j>} \mathbf{S}_{Ci} \cdot \mathbf{S}_{Cj} +J^{'\parallel}_{DD} \sum_{<i, j>} \mathbf{S}_{Di} \cdot \mathbf{S}_{Dj} \\
+&J^{\perp, 1}_{AD} \sum_{i, j} \mathbf{S}_{Ai} \cdot \mathbf{S}_{Dj} 
+J^{\perp, 1}_{BC} \sum_{i, j} \mathbf{S}_{Bi} \cdot \mathbf{S}_{Cj} \\
+&J^{\perp, 2}_{AD} \sum_{i, j} \mathbf{S}_{Ai} \cdot \mathbf{S}_{Dj} 
+J^{\perp, 2}_{BC} \sum_{i, j} \mathbf{S}_{Bi} \cdot \mathbf{S}_{Cj} \\
-&\alpha \sum_{i, j} {S}^y_{Ai} \cdot {S}^y_{Bj} +{S}^y_{Ci} \cdot {S}^y_{Dj} \\
+&\frac{\Lambda}{2}\biggl(\sum_{M}\sum_i (S^y_{Mi})^2 \biggr) - \frac{K}{2} \biggl(\sum_{M}\sum_i (S^z_{Mi})^2 \biggr) \\
-  &H_0 \biggl(\sum_{M}\sum_i S^z_{Mi} \biggr)
\end{align*}
where $M= \{A,B,C,D \}$, $J^{\parallel}_{AB}, J^{\parallel}_{CD}>0$  are the AFM nearest-neighbor in-plane exchange couplings,  and $J^{'\parallel}_{MM}<0$ are the next-nearest-neighbor in-plane exchange couplings, $J^{\perp, 1}_{AD}, J^{\perp, 1}_{CD}>0$ are the AFM out of plane coupling within the crystallographic unit cell, $J^{\perp, 2}_{AD}, J^{\perp, 2}_{CD}<0$ are the FM out of plane coupling between atoms in neighbors  crystallographic unit cell, $\Lambda, K>0$ the easy plane and uniaxial anisotropy constants, respectively, and $B_0$ is the magnetic field along $z$ ($H_0=g\mu_B B_0$). Finally, $\alpha$ describe the anisotropy of the in-plane exchange coupling.
Given that the magnetic unit cell is larger than the crystallographic one, we use the rotating frame transformation~\cite{Chernyshev2009} for each site. For instance, for A site we get:
\begin{align*}
S^{x0}_{Ai}&=  +S^x_{Ai} \\
S^{y0}_{Ai}&=  -S^y_{Ai} \\
S^{z0}_{Ai}&= -S^z_{Ai} 
\end{align*}
in the nearest-neighbor unit cells, with the $z-$direction aligned along the spin-ordering direction $a$.
We  then use the Holstein-Primakoff transformations: 
\begin{align*}
[S^-_{Ai}, S^+_{Ai}, S^z_{Ai}]&= [\sqrt {2S_A} a_i^\dag, \sqrt {2S_A} a_i, S_A - a_i^\dag a_i ] \\
[S^-_{Bj}, S^+_{Bj}, S^z_{Bj}]&= [\sqrt {2S_B} b_j, \sqrt {2S_B} b_j^\dag, -S_B + b_j^\dag b_j ]  \\
[S^-_{Ci}, S^+_{Ci}, S^z_{Ci}]&= [\sqrt {2S_C} c_i^\dag, \sqrt {2S_C} c_i, S_C - c_i^\dag c_i ] \\
[S^-_{Dj}, S^+_{Dj}, S^z_{Dj}]&= [\sqrt {2S_D} d_j, \sqrt {2S_D} d_j^\dag, -S_D + d_j^\dag d_j ] 
\end{align*}
The Hamiltonian then becomes:
\begin{align}
H&=\sum_\mathbf{q}\mathbf{x}_\mathbf{q}^\dag H_\mathbf{q} \mathbf{x}_\mathbf{q}, \qquad
H_\mathbf{q} = 
\begin{pmatrix}
A_\mathbf{q} & B_\mathbf{q} \\
B_\mathbf{q} & A_\mathbf{q}
\end{pmatrix}
- H_0 I
\end{align}
\setlength{\arraycolsep}{0pt}
\renewcommand{\arraystretch}{1.5}
with:
{\small 
\begin{align*}
&A_\mathbf{q} = \\
&\begin{pmatrix}
4 (J^{\parallel}_{AB} S_B +  \kappa^{A}_\mathbf{q} S_A) + S_DJ^{\perp}_{AD}  & -2 \sqrt{S_A S_B} \alpha \gamma^\parallel_\mathbf{q} & 0 &  -\sqrt{S_A S_D} |J^{\perp,2}_{AD}| \gamma^{\perp *}_\mathbf{\mathbf{q}} \\
-2 \sqrt{S_A S_B} \alpha \gamma^\parallel_\mathbf{q} & 4 (J^{\parallel}_{AB} S_A +  \kappa^{B}_\mathbf{q} S_B) + S_CJ^{\perp}_{BC}  & -\sqrt{S_B S_C} |J^{\perp,2}_{CB}| \gamma^{\perp *}_\mathbf{\mathbf{q}} & 0  \\
0 & -\sqrt{S_C S_B} |J^{\perp,2}_{CB}| \gamma^\perp_\mathbf{\mathbf{q}} & 4 (J^{\parallel}_{CD} S_D +  \kappa^{C}_\mathbf{q} S_C) + S_BJ^{\perp}_{BC} & -2 \sqrt{S_C S_D} \alpha \gamma^\parallel_\mathbf{q} \\
-\sqrt{S_A S_D} |J^{\perp,2}_{AD}| \gamma^\perp_\mathbf{\mathbf{q}} & 0 & -2 \sqrt{S_DS_C} \alpha\gamma^\parallel_\mathbf{q} & 4 (J^{\parallel}_{CD} S_C +  \kappa^{D}_\mathbf{q} S_D) + S_AJ^{\perp}_{AD}   
\end{pmatrix}\\
&B_\mathbf{q} = \\
&
\begin{pmatrix}
S_A\Lambda & 4(1- \frac{\alpha}{2J^\parallel_{AB} })\sqrt{S_A S_B} J^\parallel_{AB} \gamma^\parallel_\mathbf{q} & 0  &\sqrt{S_A S_D}J^{\perp,1}_{AD} \\
4(1- \frac{\alpha}{2J^\parallel_{AB} })\sqrt{S_B S_A} J^\parallel_{AB} \gamma^\parallel_\mathbf{q} & S_B\Lambda & \sqrt{S_B S_C}J^{\perp,1}_{BC} & 0 \\ 
0   & \sqrt{S_C S_B}J^{\perp,1}_{CB}  & S_C\Lambda & 4(1- \frac{\alpha}{2J^\parallel_{CD} })\sqrt{S_C S_D} J^\parallel_{CD} \gamma^\parallel_\mathbf{q} \\ 
\sqrt{S_D S_A}J^{\perp,1}_{AD}   & 0 & 4(1- \frac{\alpha}{2J^\parallel_{CD} })\sqrt{S_D S_C} J^\parallel_{CD} \gamma^\parallel_\mathbf{q} & S_D\Lambda 
\end{pmatrix}  
\end{align*}
}
with:
\begin{align*}
\mathbf{x}_\mathbf{q} &= [a_\mathbf{q}, b_\mathbf{q}, c_\mathbf{q}, d_\mathbf{q},  a^\dag_\mathbf{-q}, b\dag_\mathbf{-q}, c\dag_\mathbf{-q}, d\dag_\mathbf{-q}]\\
\gamma^\parallel_\mathbf{q} &= \frac{1}{2} [\cos(2\pi q_x)+\cos(2\pi q_y)] \\
\gamma_\mathbf{q}'&=\cos(2\pi q_x)\cos(2\pi q_y)\\
\gamma^\perp_\mathbf{q}&=e^{-i 2\pi q_z} \\ 
\kappa^{M}_\mathbf{q}&=K+\Lambda - J^{'\parallel}_{MM}(1-\gamma_\mathbf{q}') \\
J^{\perp}&=(J^{\perp,1}+|J^{\perp,2}|).
\end{align*}
The eigenvalues are obtained by diagonalization of $gH_\mathbf{q}$, with $g=[\mathbf{x}, \mathbf{x}^\dag]$. 

\subsubsection{ Case I: Mono-atomic layers  with out of plane coupling}

For this case, which describes configuration \textit{ii} is obtained for $A=B={Fe}$, $C=D={Cu}$. We also use for simplicity $ J^{\perp,1}_{AD} =J^{\perp,1}_{CD}= - J^{\perp,2}_{AD} =J^{\perp,2}_{CD}=J^{\perp}$.
At the zone center, diagonalization of $gH_\mathbf{q=0}$ gives 4 bands, with, to first order in $J^\perp/J^\parallel_{AB}$ the values (for the case of $\Lambda=K=\alpha=H_0=0$):
\begin{align*}
\omega_1&=0\\
\omega_2&= 4\sqrt{S_{Fe}S_{Cu}J^\perp J^{ab}_{Fe}}\\
\omega_3&= 4\sqrt{S_{Fe}S_{Cu}J^\perp  J^{ab}_{Cu}}\\
\omega_4&= 4\sqrt{S_{Fe}S_{Cu}J^\perp( J^{ab}_{FeFe}+ J^{ab}_{CuCu})}
\end{align*}
The gap at the BZ center is proportional to the square root of $J_\perp$.

At $(0.0, 0.0, 0.5)$ we only have two solutions:
\begin{align*}
\omega_{1,2}&= \sqrt{8 S_{Fe}S_{Cu} J^\perp \bigg(J^{ab}_{FeFe}+ J^{ab}_{CuCu} \pm  \eta  \bigg) }\\
\eta &= \sqrt{({J^{ab}_{FeFe}})^2 + ({J^{ab}_{CuCu}})^2}
\end{align*}

At the zone border and corners there is no out of plane dispersion. At $(0.25, 0.25, 0)$, $(0.25, 0.25, 0.5)$ ($p=1$)  and  $(0.5, 0, 0)$, $(0.5, 0, 0.5)$ ($p=2$) energies are:
\begin{align*}
\omega_{1,2}&= 4 S_{Fe} | J^{ab}_{Fe} - pJ_{FeFe}^{'}| \sqrt{ 1 + \frac{S_{Cu}}{S_{Fe}} \frac{J^\perp}{ J^{ab}_{Fe}-pJ_{FeFe}^{'}} } \\
\omega_{3,4}&=4 S_{Cu} | J^{ab}_{Cu} - pJ_{CuCu}^{'}| \sqrt{ 1 + \frac{S_{Fe}}{S_{Cu}} \frac{J^\perp}{ J^{ab}_{Cu}-pJ_{CuCu}^{'}} }.
\end{align*}

\subsubsection{ Case II: Di-atomic checkerboard layers  with out of plane coupling}

The dispersion for configuration \textit{v} is obtained with $A=C={Fe}$, $B=D={Cu}$.
For this case the magnon spectrum is gapped at the zone center even for $J_\perp=0$, and the energies at the zone center are (valid to the leading order in $J_\perp/J_\parallel$, and with $dS > 0$ and $\Delta=K=\alpha=H_0=0$)
\begin{align*}
\omega_1&=0\\
\omega_2&= 4\sqrt{2}\frac{S_{Fe}S_{Cu}}{dS} J_\perp \\
% \omega_3&= 4\sqrt{ dS^2 (J^{ab}_{CuFe})^2+(dS^2+S_{Fe}S_{Cu})J^{ab}_{CuFe}  J_\perp}\\
% \omega_4&= 4\sqrt{ dS^2 (J^{ab}_{CuFe})^2+(dS^2+3S_{Fe}S_{Cu})J^{ab}_{CuFe}  J_\perp}
\omega_3&= 4 dS J^{ab}_{CuFe}\sqrt{ 1+\Big( 1+ \frac{S_{Fe}S_{Cu}}{dS^2}\Big) \frac{J^\perp}{J^{ab}_{CuFe}} }\\
\omega_4&= 4 dS J^{ab}_{CuFe}\sqrt{ 1+\Big( 1+ \frac{3S_{Fe}S_{Cu}}{dS^2}\Big) \frac{J^\perp}{J^{ab}_{CuFe}} }
\end{align*}

At (0,0, 0.5) $\omega_3$ and $\omega_4$ are the same as at $\Gamma$ and:
\begin{align*}
\omega_1&= 2\sqrt{4-2\sqrt{3}}\frac{S_{Fe}S_{Cu}}{dS} J_\perp \\
\omega_2&= 2\sqrt{4+2\sqrt{3}}\frac{S_{Fe}S_{Cu}}{dS} J_\perp 
% \omega_3&= 4\sqrt{ dS^2 (J^{ab}_{CuFe})^2+(dS^2+S_{Fe}S_{Cu})J^{ab}_{CuFe}  J_\perp}\\
% \omega_4&= 4\sqrt{ dS^2 (J^{ab}_{CuFe})^2+(dS^2+3S_{Fe}S_{Cu})J^{ab}_{CuFe}  J_\perp}
\end{align*}

At $(0.25, 0.25, 0)$, $(0.25, 0.25, 0.5)$ ($p=1$)  and  $(0.5, 0, 0)$, $(0.5, 0, 0.5)$ ($p=2$) the energies are:
\begin{align*}
\omega_{1,2}&=4 |J^{ab}_{CuFe} - pJ_{FeFe}^{'}| S_{Cu} \sqrt{1+\frac{J_\perp}{J^{ab}_{CuFe} - pJ_{FeFe}^{'}}}\\
\omega_{3,4}&=4 |J^{ab}_{CuFe} - pJ_{CuCu}^{'}| S_{Fe} \sqrt{1+\frac{J_\perp}{J^{ab}_{CuFe} -  pJ_{CuCu}^{'}}}
\end{align*}

\section{LSWT in YBCFO: corrections due to quantum fluctuations and higher orders. } \label{ring}

It is well established that in cuprates the spin-wave excitations of the Cu $d^9$ spins are affected by both quantum fluctuations and ring exchange coupling $J_c$. 
The main effect of quantum fluctuation can be included by a renormalization of the spin-wave dispersion by a factor $Z_c\approx 1.18$~\cite{Singh1989}. The effect of ring exchange coupling is equivalent to  an independent renormalization of  in-plane $J^{eff}_{NN}$ and $J'^{eff}_{NNN}$~\cite{Coldea2001}. 
In La$_2$CuO$_4$ (LCO) $J^{eff}_{NN}$ is decreased by ring exchange by approximately $14\%$, which combined with quantum fluctuation correction gives a negligible combined renormalization of $J^{eff}_{NN}$~\cite{Katanin2002} with respect to the DFT value. So in our LSWT calculations we consider $J^{eff}_{NN}\approx J^{DFT}_{NN}$.

The effect of ring exchange correction is more dramatic for $J{'}_{NNN}$. DFT calculations in LCO predict an AFM NNN-coupling $\sim 16$ meV~\cite{Annett1989}, while experiments show an effective FM coupling $\sim -10$ meV~\cite{Coldea2001}. Similarly, our DFT calculation for diatomic planes results in an AFM coupling of about $24.53$ meV for Cu atoms, while the experiments show the dispersion corresponding to an effective FM exchange. Following the relations in~\cite{Coldea2001, Katanin2002}, based on the $t-J$ model, we can estimate the effective NNN coupling using $J^{eff}_{NN}$ as:
\begin{align*}
    J'^{eff}_{NNN} = Z_c(J'-J_c/4) \approx \Bigg(\frac{16}{4}-\frac{1}{4}\bigg(\frac{U}{t}\bigg)^2 \Bigg)^{-1}J^{eff}_{NN}.
\end{align*}
The value of $U/t$ can be obtained by asking $J^{eff}_{NN}\approx 4t^2/U-24 t^4/U\approx 130$ meV, the DFT value in YBCFO. Fixing $U=3$ we have $U/t\approx9$ and we get $J'^{eff}_{NNN}=-8$ meV, which we use in the calculations of Fig.~2 in the main text.

\section{Damping factor dependence in ASD calculations}

\begin{figure}[t]
    \centering
    \includegraphics[width=1\columnwidth]{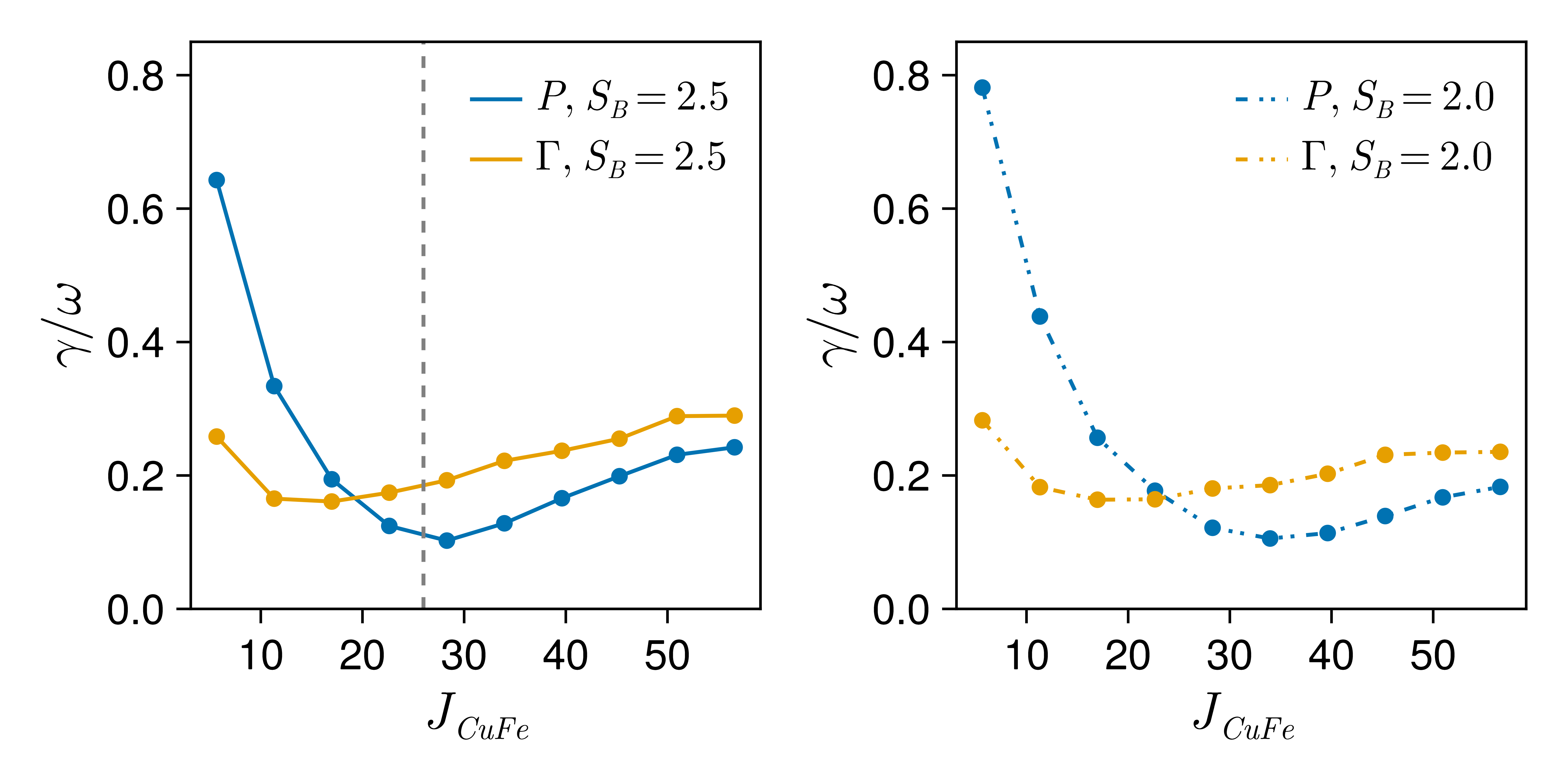}
    \caption[JCuFe_dep]{Calculation of the damping factor for the optical branch of the maximal entropy phase as function of the $J_{CuFe}$ for Fe$3+$ and Fe$2+$. Dashed line in  Fe$3+$ represent the $J^{ab}_{CuFe}=26$ meV, we obtained from our RIXS data for YBCFO.}
    \label{JCuFe_dep}
\end{figure}

The calculated damping factor for the optical branch in the maximal entropy phase as a function of the value of $J_{CuFe}$ is shown in Fig.~\ref{JCuFe_dep}~($J^{ab}_{CuCu}=125$, $J^{ab}_{FeFe}=9$ meV).  The calculation confirms the expectation for smallest  $\gamma/\omega$ at $P$ for an invariant local field, with $J^{ab}_{CuFe}=J^{ab}_{CuCu}S_{Cu}/S_{Fe}=25, 31.25$ meV, for $S_{Fe}=2.5, 2.0$ respectively.

\section{Effect of spiral order on magnon dispersion } \label{AppendixSpiral}

\begin{figure*}[t]
    \centering
    \includegraphics[scale=.35]{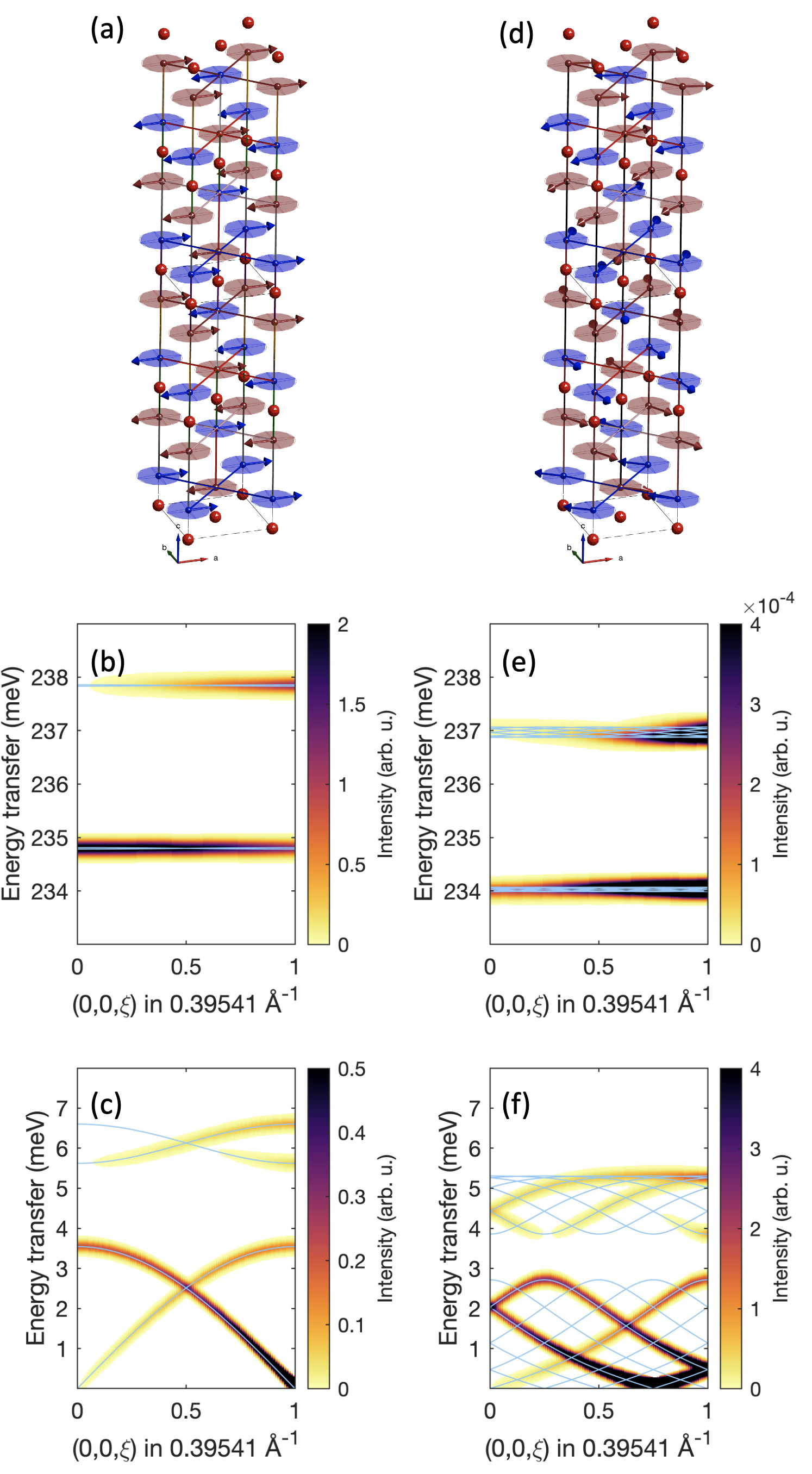}
    \caption[Fig2]{ Magnetic structure and LSWT calculations for commensurate (a)-(c) and spiral (d)-(f) ground state magnetic ordering. In (a) Spin length is renormalized to 1 for better visualization. The shaded disc represent a small easy plane anisotropy ($\Lambda=0.5$ meV) used in the LSWT calculations. In (b)-(c) and (e)-(f), calculation are performed along the out of plane  direction $ \mathbf{q} = (0 , 0, \xi)$. Continuous line are the magnons eigenenergies $\omega(0, 0, \xi)$. Intensity plots represent the real part of the trace of the spin-spin
    correlation $\mathscr{S}(\mathbf{q},\omega)$.  }
    \label{FigSpiral}
\end{figure*}

To estimate the changes in the magnon dispersion between the CM and spiral phase of YBCFO, we  approximated the average effect of the impurities by introducing a weak ferromagnetic next nearest neighbor (NNN) interaction along c at Fe sites in the pure system. 
For simplicity we consider only configuration \textit{v} of Fig.~\ref{FigS1}.
A value of $J^{c}_{Fe-FeNNN} = -0.16$ meV is used to obtain a spiral order with $q_S=0.125$ (1/8), for a single domain with ($q_S>0$), while $J^{c}_{Fe-FeNNN} = 0$ meV for the commensurate calculations. Calculations are performed using SpinW~\cite{SpinW}.
The effect of $J^{c}_{Fe-FeNNN}\neq 0$ is twofold. First, a small shift of about $1.5-2$ meV is evident in Fig.~\ref{FigSpiral}. The shift is only present in $T=0$ calculations and should not be expected in a high temperature CM phase.   Second, and more relevant, the stabilized spiral order has a $2/q_S=4$ larger magnetic cell with respect the CM phase along $c$. Thus, we expect a 4-fold increase in the number of the spin bands for the dispersion along $\mathbf{q}_c$, i.e. in $\omega(0,0,\xi)$, which is also observed in (e) and (f). 
In summary, we see that the effect of the ICM is negligible, especially at Cu [panels (b) and (e)], and it has the largest (but still small) effect for $\mathbf{q}$ out of plane.
The almost negligible spin dispersion along c also confirms the quasi-2D nature of the magnons in YBCFO, as expected from the small values of $J$ perpendicular to the $ab$ plane.

\end{document}